\documentclass[12pt,preprint]{aastex}
\newcommand{\be}{\begin{equation}}
\newcommand{\ee}{\end{equation}}
\newcommand{\bea}{\begin{eqnarray}}
\newcommand{\eea}{\end{eqnarray}}

\begin{document}
\title{Optical polarization from aligned atoms as a new diagnostic of 
interstellar and circumstellar magnetic fields}
\author{Huirong Yan \& A. Lazarian}
\affil{Department of Astronomy, University of Wisconsin, 475 N. Charter
St., Madison, WI 53706; yan, lazarian@astro.wisc.edu}

\begin{abstract}
Population of levels of the hyperfine and fine 
split ground state of an atom is affected by radiative transitions induced 
by anisotropic radiation flux. 
Such aligned atoms precess in the external 
magnetic field and this affects properties of polarized radiation arising 
from both scattering and absorption by atoms. As the result the degree of 
light polarization depends on the direction of the magnetic field. This 
provides a perspective tool for studies of astrophysical magnetic fields using
optical and UV polarimetry. We provide calculations for several atoms
and ions that can be used to study 
magnetic fields in interplanetary medium, interstellar medium,
circumstellar regions and quasars.

\end{abstract}

\section{Introduction}

Magnetic fields play extremely important role in Astrophysics.
Polarimetry of aligned dust provides one of the ways of studying magnetic
field (see review by Lazarian 2003). Polarimetry of some molecular line
 have been recently shown to be a good tool for magnetic field  studies (see
Girart, Crutcher \& Rao 1999). 
 Here we 
discuss yet another promising technique to study magnetic fields that
employs
optical and UV polarimetry. The proposed technique is based on
the ability of atoms to be aligned by external radiation in their ground state
and be realigned through precession in magnetic  field. 

It has been known that  
atoms can be aligned through the interactions with the anisotropic flux
of resonance emission (see review by Happer 1972 and references therein). Alignment is  understood here in terms of orientation
of vector angular momentum
$\bf J$, if we use the language of classical mechanics. In quantum
terms  this means a difference in the population of sublevels corresponding to
the projections of angular momentum to the quantization axis. We will
call atomic alignment only the alignment of $\bf J$ in the ground state.
This state is long lived and therefore is being affected by
weak magnetic field. This is the effect that we are dealing with below.

Atomic alignment has been studied in laboratory in relation with early day maser 
research (see Hawkins 1955). Alignment of atoms in ground state can change
the optical properties of the medium.  This effect was noticed and made
use of by Varshalovitch (1968) in case of hyperfine structure of the ground state, for 
fine structure of the ground state by, e.g. Lee,  Blandford \& Western (1994).

In this paper we present a discussion of the atomic alignment in the presence of 
magnetic field and outline the features of a new perspective technique for
studies of astrophysical weak magnetic fields.  Indeed,
magnetic field of rather small amplitude
is able to induce precession of aligned atoms and thus influence the 
emanating polarization. Astrophysical magnetic fields are frequently too weak to cause a substantial Zeeman (see Heiles 1997, Girart, Crutcher \& Rao 1999) or Hanle (see Nordsieck 1999) effect.
Therefore it is really advantageous to use this technique.
Moreover, we shall show that it provides information that is  not available
using any other tool. The possiblity of using atomic alignment to study weak magnetic field in diffuse media was first discussed in Landolfi \& Landi Edgl'Innocenti (1986) for two-level atoms with fine structures and polarization of emissions from these atoms  were considered for two opposite magnetic inclinations.

Below we show that atomic alignment  is present for a variety of atoms and we aim at obtaining the one-to-one correspondence between polarizations of both emission and absorption lines and direction of magnetic field for various atoms. In particular, we discuss alignment of atoms with fine and hyperfine splitting
of ground state levels. 
 NaI, NV, KI,  HI, AlIII, OI, OII, NI, CrII, C II, OIV and CI, O III are presented
as examples of a broad variety of
atomic species that can be aligned by radiation. This
opens wide avenues for the magnetic field research in circumstellar regions,
interstellar medium, interplanetary medium, intracluster medium and quasars etc. More details about practical limitations of the technique will be given in a companion paper, while here we concentrate on discussing basic physics of the
phenomenon of atomic alignment and the  influence of magnetic field. 

The discussion of atomic alignment physics is presented in \S2 both using
toy models and quantum mechanics calculations. The relation between the
polarization of scattered radiation and atomic alignment is given in
\S3, while similar relation between atomic alignment and adsorbed radiation
is discussed in \S4. Earlier work is briefly mentioned in \S5. The discussion
and the summary are provided in, respectively, \S6  and \S7.  

\section{Alignment physics: toy model}

Radiation from stars and other astrophysical sources largely affect
the surrounding medium.  While the aspect concerning
momentum have been studied extensively, i.e., stellar wind, outflow
and jets, radiation force on dust, Eddington limit, etc., the study
of the effects involving angular momentum are far behind.

The basic idea of the atomic alignment is simple. The alignment is caused by
the anisotropic deposition of angular momentum from photons. In typical
 astrophysical situations the radiation
flux is anisotropic. As the photon
spin is along the direction of its propagation, we expect that atoms
that scattered the radiation can be  aligned
in terms of their angular momentum. Such an alignment happens in terms of 
the projection of angular momentum
to the direction of the incoming light. It  is clear that to have the
alignment of the ground state, the atom should have non-zero angular
momentum in its ground state. Therefore fine or hyperfine structure\footnote{
A natural question is why one should consider the {\it  total} angular
momentum of the atom, i.e. consider  hyperfine splitting, while 
the nuclei moment does not change in optical transitions. 
The answer, which has not been frequently appreciated in spectroscopic
literature is that the ground state that we consider is long-lived compared
with the rate of the precession of electrons in the magnetic field
of the nuclei. As the result, coupling of nuclei and electrons takes
place.}
of the ground state is necessary to enable various projection of
angular momentum to exist in the ground state.

Before we present detailed calculations, let us discuss a couple of
toy models that provide an intuitive  insight into  the physics 
that we deal with here.
First of all, consider a toy model of an
atom with lower state corresponding to the total angular momentum
$I=1$ and the upper state corresponding to angular momentum $I=0$.
If the projection of the angular momentum to the direction of the
incident resonance photon beam is $M$, for the lower state $M$ can
be $-1$, $0$, and $1$ , while for the upper state $M=0$ (see Fig.1). 
The unpolarized
beam contains an equal number of left and right circularly polarized
photons which projection on the beam direction are 1 and -1. Thus
absorption of the photons will induce transitions from $M=-1$ and
$M=1$ states. However, the decay of the upper state happens to all
three levels. As the result the atoms get accumulated at the $M=0$ ground
state from which no excitations are possible. As a result of that the optical
properties of the media (e.g. absorption) would change\footnote{
Not every type of alignment affects  the polarization of the
scattered  of absorbed  radiation. Interestingly
enough,  alignment that we discuss within the toy model
does not induce any extra
polarization for emissions. To have emission polarized, the alignment on the ground state should be transfered to the uneven occupation on the excited level. Therefore atoms with more complex structure
of the excited levels should be considered.}.

This above toy model can also exemplify the
 role of collisions and magnetic field. Without collisions one may expect 
that all atoms
reside eventually at the state of $M=0$. Collisions, however, redistribute
atoms with different states. However, as disalignment of the ground
state requires spin flips, it is less efficient than one can naively
imagine. Calculations by Hawkins (1955) show that to disalign sodium
one requires more than 10 collisions with paramagnetic atoms and experimental
data by Kastler (1956) support this. This reduced sensitivity of aligned
atoms to disorienting collisions makes the effect important for various
astrophysical environments.

\begin{figure}
\plottwo{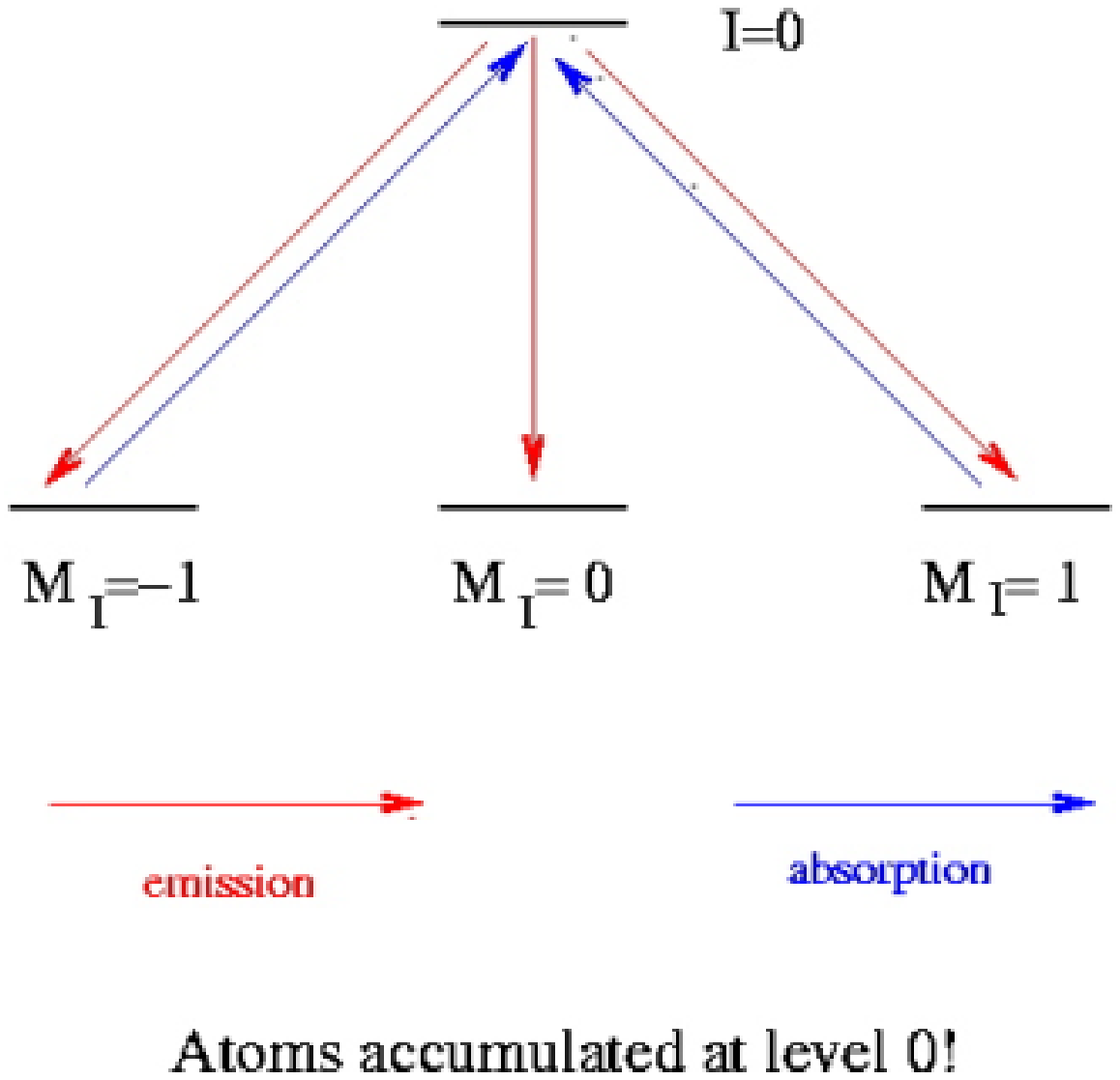}{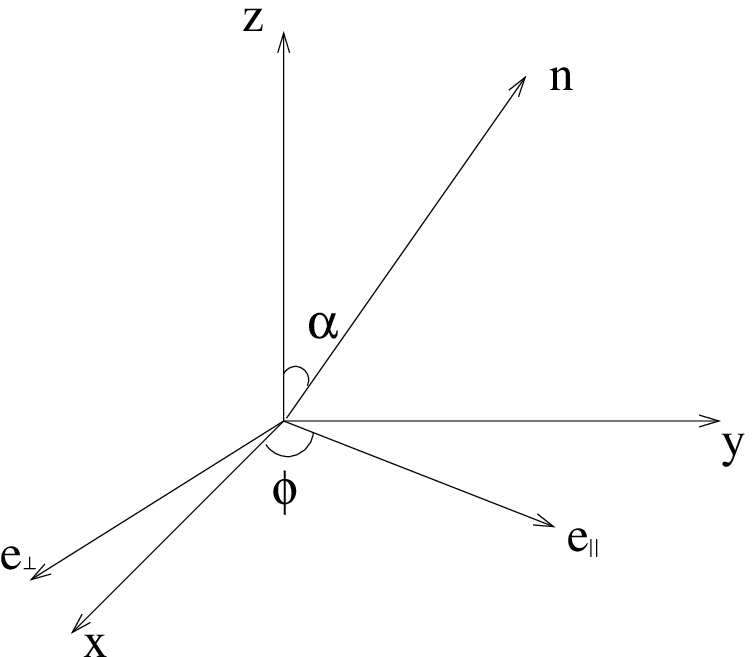}
\caption{(a) A toy model to illustrate how atoms are aligned by anisotropic light.
Atoms get accumulated in the ground state $M_I=0$, as radiation removes atoms from the ground states $M_I=1$ and $M_I=-1$; (b) Geometry of the radiation system. ${\bf z}$ is defined by the direction of incident light, and thus it is the alignment axis. {\bf n} refers to the the direction of emitted or absorbed light (see text). }
\end{figure}

Magnetic field would also mixes up different $M$ states. However, it
is clear that the randomization in this situation will not be complete
and the residual alignment would reflect the magnetic field direction
in respect to the observer. Magnetic mixing happens if the angular
momentum precession rate  is higher than the rate of the
excitation of atoms from the ground state, which is true
for many astrophysical conditions when we deal with
weak magnetic fields. Therefore,  further on we discuss the weak field
limit when the Zeeman splitting of
the upper level of the atom is much smaller than the energy width of the 
upper level.  

Note that in order to be aligned, first the atoms should have enough 
degree of freedom, namely, the quantum number of angular momentum must be 
$\ge 1$\footnote{With total angular momentum $=1/2$, atoms can only be 
oriented by polarized light.}. Second, incident flux must be anisotropic. 
Moreover, the collisional rate should not be too high. While the latter
requires special laboratory conditions, it is the
case for many astrophysical environments such as the outer
layers of stellar atmosphere, interplanetary,
interstellar, and intergalactic media, etc.

As long as these conditions are satisfied, atoms can be aligned within either 
fine or hyperfine structure. For light elements,
the hyperfine splitting is very small and the line components overlap
to a large extent. However, for resonant lines, the hyperfine interactions causes 
substantial 
precession of electron angular momentum ${\bf J}$ about
the total angular momentum ${\bf F}$ before spontaneous emission. Therefore 
total angular momentum should be considered and the $FM_{F}$
basis must be adopted (Walkup et al. 1982). For alkali-like atoms,
 hyperfine structure should
be invoked to allow enough degrees of freedom to harbor the alignment and to 
induce the corresponding polarizations.

In terms of time scales, we have a number of those, which makes the problem
interesting and allows getting additional information about environments.
The corresponding rates 
are 1) the rate of the precession $\tau_L^{-1}$, 2) the rate of the photon
arrival $\tau_A^{-1}$, 3) the rate of collisional randomization $\tau_R^{-1}$,
4) the rate of the transition within fine or hyperfine structure $\tau^{-1}_T$.
For the sake of simplicity in the bulk of the paper
we shall discuss the situation that
$\tau_L^{-1}>\tau_A^{-1}>\tau_R^{-1}>\tau_T^{-1}$. Other relations are possible, however.
If the Larmor precession gets comparable with any of the other rates,
it is possible to get information about the {\it magnitude} of magnetic
field. Another limitation of our approach is that we consider that $\tau_L^{-1}$
is much smaller that the rate of the decay of the excited state, which
means that we disregard the Hanle effect.

\subsection{Optical pumping }

A beam of photons creates optical pumping of atom's ground state.
Transition probability from the initial state $n$ to the final state  $m$
can be obtained 
according to a standard, time-dependent perturbation theory, 
the amplitude of resonant scattering is (Stenflo 1999)

\be
S_{0m}=\Sigma_n\frac{<m|V^i_q|n><n|V^o_q|0>}{\omega_{nm}-\omega-{i\gamma/2}},
\ee
where the summation over $n$ represents different scattering routes through different excited states permitted by selection rules, $V_q^{i,o}={\bf r\cdot e}_q^{i,o}$ is the projection of dipole moment 
along basis vector ${\bf e}_q$ of incoming and outcoming light,  
${\bf e_{\pm}}= (\mp{\bf \hat x}-{\bf \hat y})/\sqrt{2}$, 
${\bf e}_0={\bf \hat z}$. The standard notations for the frequencies $\omega$
and $\omega_{nm}$ as well as for the damping rate of the excited state
$\gamma$ are employed.

According to Wigner-Eckart theorem, the 
amplitude of transition from a {\it fine} level $J$ to $J'$ for electric 
dipole radiation along ${\bf e}_q$ (Cowan 1981)
\begin{eqnarray}
R^q_{JJ'}&=&<JM| V_{q} |J'M'>\nonumber\\
&=&(2J+1)^{-1/2}C(J'M'q;JM)<J||V||J'>,
\label{fine}
\end{eqnarray}
where $C$ are the Clebsch-Gordan coefficients (see Condon \& Shortley 1951)
and $<J||V||J'>$ is the square root of line strength, which scales out for  
 the calculation of polarization.  The corresponding absorption 
from $J'$ to $J$ will have $R_{J'J}=R_{JJ'}^{*}$, where $^{*}$ denotes the
conjugate quantity.

For {\it hyperfine} lines, in the case of weak interaction in which 
neighboring fine levels do not interact, so that $J$ and $J'$ are good 
quantum numbers. The amplitude of the components of a hyperfine structure 
multiplet is then given by

\begin{eqnarray}
&R^q_{FF'}&=<IJFM_F|V_q|IJ'F'M'_F>=C'(IJMM_J;FM_F)\nonumber\\
&&<JMM_J|V_q|J'M'M'_J>C(IJ'M'M'_J;F'M'_F),
\label{hypfine}
\end{eqnarray}
where $M$ and $M_J$ are quantum numbers corresponding to 
the projection of the total angular momentum $F$ and
its electron part $J$ respectively, $I$ corresponds to the angular momentum
of the nuclei. 
The matrix above, $<JMM_J|V_q|J'M'M'_J>=<JM_J|V_q|J'M'_J>$,
 doesn't depend on the $M$ because the direct interaction between the nucleus 
and the radiation field is negligible. 

In general, the incident radiation is not unidirectional. For this case we can 
define the symmetry axis of the radiation as the quantization axis 
${\bf \hat{z}}$. If the vectors parallel and perpendicular to the 
${\bf n-\hat{z}}$ plane (see Fig. 1), ${\bf e_\parallel}$ and ${\bf e_\perp}$, 
are adopted as the basis 
vectors for polarization, then for the light absorbed or emitted in direction 
${\bf n}(\alpha, \phi)$, the matrix element for the transition is 
(Lee \& Blandford 1997):

\begin{eqnarray}
R({\bf e}_\parallel)&=&[2^{-1/2}\cos\alpha e^{i\phi}R^{-1}, -\sin\alpha R^0, -2^{-1/2}\cos\alpha e^{-i\phi}R^1]\nonumber\\
R({\bf e}_\perp)&=&[2^{-1/2}ie^{i\phi}R^{-1}, 0, 2^{-1/2}ie^{-i\phi}R^1],
\label{decomp}
\end{eqnarray}
 
After a single scattering event, the occupation on the ground state will be changed according to Eq.(1)

\be
\label{scatter}
|S_{0m}|^2=\frac{1}{\gamma}\Sigma_n |R_{0n}|^2|R_{nm}|^2+i\frac{\Sigma_{nn'}R_{mn}R_{n0}R^*_{0n'}R^*_{n'm}}{\omega_{nn'}-{\it i}\gamma},
\ee
where $\gamma$ is the natural linewidth, $\omega_{nn'}$ refers to the frequency difference between 
two excited sublevels $n$ and $n'$. The second term on the right hand arises from coherence between $n$ and $n'$. 
It becomes important only when the coherence frequency 
does not differ too much from the natural width, e.g., $\omega_{nn'}\gtrsim\gamma$.

\subsection{Mixing by magnetic fields}

We have discussed in the previous section that atoms can be aligned by
radiation. The effect of magnetic field on the aligned atoms is to change
the alignment through the Larmor precession. As the result the alignment
gets the imprint of magnetic field orientation. 

As we mentioned earlier in this paper we consider the case when the photon arrival 
rate $\tau_A^{-1}$ is much smaller
than the Larmor precession rate $\tau_L^{-1}$. For optical lines, the life time of the excited 
state can be ignored comparing
with the Larmor precession if the field is less than $\sim1$ Gauss Hawkins (1954, hereafter H54). 
Thus the atom immediately after scattering can be described with respect to z axis as if there were 
no magnetic field.  The subsequent effect of magnetic field is mainly embodied  in the mixing within 
the ground state before the atom is excited by another photon. The mixing due to the Larmor 
precession can be described by the variation of the angular momentum in Heisenberg representation 
in which the state functions do not change (H54). The precession of axial angular momentum 
is then similar to the classical case: 
\begin{eqnarray}
F_{z}(t) & = & F_{z}(0)[\cos^{2}\theta+\sin^{2}\theta\cos\Omega t]+
F_{x}(0)\sin\theta\cos\theta\nonumber\\
& &(1-\cos\Omega t) + F_{y}(0)\sin\theta\sin\Omega t,\label{Fzt}
\end{eqnarray}
where $F_{x},F_{y},F_{z}$ are the projections of the angular momentum
along $x,y,z$ directions, $t=0$ represents the time right after
one scattering event, $\theta$ is the angle between $z$ and magnetic
field. The variation of the population on ground sublevels will be
entirely reflected by the time-dependant projection operator $P_{m}$
since the eigenstates $\psi_{m}$ are invariables. The projection
operator defined as $P^f_{m}\psi_{m'}=\delta_{mm'}\psi_{m}$ can be
expressed as a function of $F_{z}$: 
\begin{equation}
P^f_{m}=\frac{(F_{z}/\hbar+f)}{(m+f)}\frac{(F_{z}/\hbar+f-1)}{(m+f-1)}\cdots\frac{(F_{z}/\hbar-f)}{(m-f)},\label{proj}\end{equation}
where $f$ is the quantum number of total angular momentum and
$m$ is that of the axial angular momentum $F_{z}$. The expression is constructed in such a way that 
only when the axial angular momentum is equal to $m\hbar$ the projection gives non-zeros result. 
This is because for a total angular momentum f, the axial angular momentum $F_z$ can take only
the values $f\hbar, f-1\hbar\cdots -f\hbar$. Therefore the numerator of $P^f_m$ is always zero.

In the representation where the basis vectors are the eigenstates of $F_z(0)$, 
the angular momentum operators are given by (Sobelman 1972): 

\begin{eqnarray}
<fm+1|F_x(0)|fm>&=&1/2\sqrt{(f-m)(f+m+1)}, \nonumber\\
<fm-1|F_x(0)|fm>&=&1/2\sqrt{(f+m)(f-m+1)}, \nonumber\\
<fm+1|F_y(0)|fm>&=&-i/2\sqrt{(f-m)(f+m+1)}, \nonumber\\
<fm-1|F_y(0)|fm>&=&i/2\sqrt{(f+m)(f-m+1)}.
\end{eqnarray}

The examples of the appropriate matrixes are given in the Appendix A.

Since the precession rate is much higher than the rate of 
absorption events, the ensemble averaged observables results
can be obtained by taking the time average. Combining Eqs.(\ref{Fzt})
and (\ref{proj}), we can get the time-averaged projection operator $\overline{P}_{m}(\theta)$. 
The ground population averaged by the precession around the magnetic field can be then represented by 
 the expectation value of the projection operator (Merzbacher 1970):

\begin{equation}
\rho'=<\overline{P}_{m}(\theta)>=trace(\rho \overline P_m).
\label{precess}
\end{equation}
If density matrix is diagonal, we can use an occupation vector $(\rho_1, \rho_2, ...)$ to represent 
the ground population. 
In that case, the magnetic mixing will be simply represented by a transformation matrix:

\be
B_{mm'}=\psi^*_m \overline P_m' \psi_m.
\ee
Apparently the mixing due to the precession increases with the angle $\theta$ between the 
magnetic field and z-axis. Atoms obtain the maximum degree of alignment when 
the magnetic field is parallel to z axis and 
no mixing is present. The occupation in this case is simply the result of 
optical pumping.
When the magnetic field is perpendicular to the magnetic field, the occupation
is to a large degree uniform. Fig.2 shows the time-averaged occupation on 
ground sublevels
 for $\theta=0$ and $\theta=\pi/2$ for NaI and NV. 

\begin{figure}
\plotone{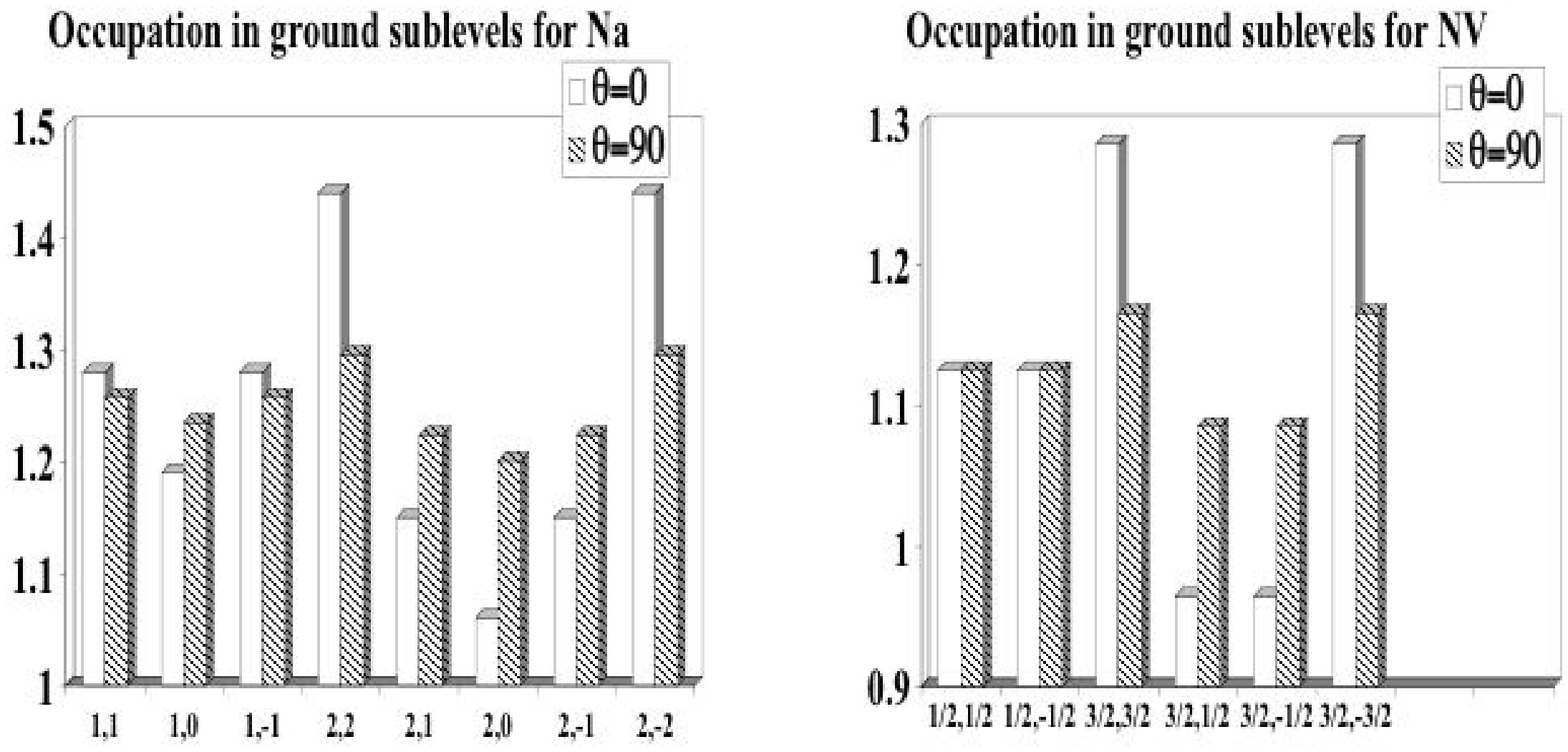}
\caption{Differential occupations among hyperfine sublevels of 
NaI and NV ground states for two different magnetic configurations (see text). }
\end{figure}

\section{Polarization of Scattered Light}

\subsection{Alignment within hyperfine structure}

Alignment of NaI which has a hyperfine splitting of the ground state has 
been experimentally
studied by Hawking (1955). The alignment was proved to be efficient with 
unpolarized light.

{\bf NV and NaI}\\

NV and Na I have similar electron configurations. We 
shall discuss NV. The ground state of NV is $2^{2}S_{1/2}$ and the first
excited states $2^{2}P_{1/2}$, $2^{2}P_{3/2}$ correspond to
D1 and D2 lines respectively. 
The nuclear spin of NV is $I=1$. The total angular
momentum of the ground state thus can be $F=(1\pm1/2)\hbar=3\hbar/2,\hbar/2$.
Therefore the ground state has totally $(2\times3/2+1)+(2\times1/2+1)=6$
sublevels which enables alignment. 
For NV, the hyperfine splitting 
is much larger than the natural width of the ground state. On ground level, the smallest hyperfine 
splitting is about $5.6\gamma$. Since it is the square of frequency difference that is weighed, we 
can safely ignore the interference term. In this case the density matrices of both the ground and 
excited states will be diagonal.  

Suppose that originally the ground sublevels are equally populated. 
In what follows, we shall limit our discussions to the case when the incident light is coming 
from ${\bf \hat z}$ direction. So there will only be excitations with 
$\delta F_z=\pm 1$. Then after a single 
scattering event, the ground population according to 
eqs.(\ref{decomp}) and (\ref{scatter}) without the interference term will be

\begin{eqnarray}
\label{newrho}
\rho_{m}&=&\Sigma_{0} \rho_0 |S_{0m}|^2\nonumber\\
&=&\frac{1}{\gamma}\Sigma_{0,n}\rho_0|R_{0n}|^2 |R_{nm}|^2\nonumber\\
&=&\frac{1}{\gamma} \Sigma_{0,n} \rho_0 (R^1_{0n}+R^{-1}_{0n})^2((R_{nm}^1+R_{nm}^{-1})(1+<\cos^2\alpha>)/2+(R_{nm}^0)^2<\sin^2\alpha>),
\end{eqnarray}
where $\alpha$ is the scattering angle, over which the above expression is averaged. Using 
Eq.(\ref{hypfine}), we can obtain the scattering matrix for both D1 and D2 lines. The scattering 
probability $|S_{0m}|^2$ for D2 line is presented in Table 1. Similarly, the emissivity from 
atoms with a ground population $\propto \rho_g$ will be:

\begin{eqnarray}
j_{\parallel}&=&\Sigma_{0,n} \rho_g (R^1_{0n}+R^{-1}_{0n})^2((R_{nm}^1+R_{nm}^{-1})\cos\alpha/2+R_{nm}^0\sin\alpha)^2,\nonumber\\
j_{\perp}&=&\Sigma_{0,n} \rho_0 (R^1_{0n}+R^{-1}_{0n})^2(R_{nm}^1+R_{nm}^{-1})^2/2.
\label{emissivity}
\end{eqnarray}

\begin{deluxetable}{lcccccc}
\tablecaption{The scattering probability $|S_{gg'}|^2$ of NV D2 line due to pumping by unidirectional unpolarized light (see Eq.(\ref{newrho}) and text around).}
\tablewidth{0pt}
\tablehead{
\colhead{~~~~~$FM_F$} & \colhead{$\frac{3}{2},\frac{3}{2}$} & \colhead{$\frac{3}{2},\frac{1}{2}$} &
\colhead{$\frac{1}{2},\frac{1}{2}$} & \colhead{$\frac{1}{2},-\frac{1}{2}$} &\colhead{$\frac{3}{2},-\frac{1}{2}$}& \colhead{$\frac{3}{2},-\frac{3}{2}$} \\
\colhead{$FM_F$} &  & & & & &}
\startdata
$\frac{3}{2},\frac{3}{2}$&
0.588&
0.0379&
0.0463&
0.037&
 0.0412&
0\\ \hline
$\frac{3}{2},\frac{1}{2}$&
0.162&
0.303&
0.0864&
0.0525& 
0.106&
0.0412\\ \hline
$\frac{1}{2},\frac{1}{2}$&
0.0833&
0.0864&
0.39&
0.137& 
0.0154&
0.037\\ \hline
$\frac{1}{2},-\frac{1}{2}$&
0.037&
0.0154&
0.137&
0.39&
0.0864&
0.0833\\ \hline
$\frac{3}{2},-\frac{1}{2}$&
0.0412&
0.106&
0.0525&
0.0864& 
0.303&
0.162\\ \hline
$\frac{3}{2},-\frac{3}{2}$& 
0&
0.0412&
0.037&
0.0463&
0.0379&
0.588\\ 
\enddata
\end{deluxetable}

If there exists magnetic field, the precession around the field will repopulate 
ground sublevels according to Eq.(\ref{precess}). If there are multiple scattering events, 
the density matrix of the ground state should be multiplied  by $S_{gg'}$ and $B_{gg'}$.

In optically thin case, the linear polarization will be
\be
P=\frac{j_\parallel-j_\perp}{j_\parallel+j_\perp},
\label{scatterpol}
\ee
We shall calculate polarization using polarizabilities (Stenflo 1994).
The corresponding polarizability can be obtained from the linear polarization of emission at 
$\alpha=90^o$, e.g.,  $E_1=4P(90^o)/(P(90^o)+3)$, where $P(90^o)$ is obtained from 
Eqs.(\ref{emissivity}) and (\ref{scatterpol}).
The polarization for light scattered to other angle is then simply given by

\be
P=\frac{3E_1\sin^2\alpha}{4-E_1+3E_1\cos^2\alpha}.
\label{polarization}
\ee  
 
We calculated the polarization for D2 line (see Fig.\ref{E1.eps}). 
In general, transitions from state with less ($2J_i+1$) number sublevels to state with more or equal 
($2J_f+1$) sublevels are 
 unpolarized or marginally polarized unless hyperfine lines can be resolved (see Fig.8 and \S 4). This is because the probability of different transitions becomes 
equal or comparable. Particularly, transitions of $\Delta J=1$ are unpolarized if the final 
(no matter ground or excited) $J_f=1$, or 3/2; for $\Delta J=0$, transitions with $J_f=1/2$ are 
unpolarized. 
It should be noted that D1 transition should be taken into account in order to get the right 
polarization for D2 line because all the transitions affect the ground population. This is crucial 
to obtain correct polarization degree for more complicated multiplets.

All the calculations in this paper are based on the general routine described above.
In practical terms calculations were performed using MatLab. First, we calculate the transition 
probabilities $R$ and construct the scattering matrix $S_{0m}$ (Eqs.\ref{scatter} and \ref{newrho}). 
Depending on the quantum number, the magnetic mixing matrix $B_{mm'}$ is then acquired according the
 routine described in \S2.2. Then the ground population can be obtained by multiplying $S_{0m}$ and 
$B_{mm'}$ 
sequentially depending on the number of scattering events. With the quantified alignment known, we
 finally can get the emissivity and polarization. 
The difference between different species comes primarily from their different structures and 
transition probabilities. All the lines and corresponding atomic properties are listed in Table2.

\begin{figure}
\label{Nasc}
\plottwo{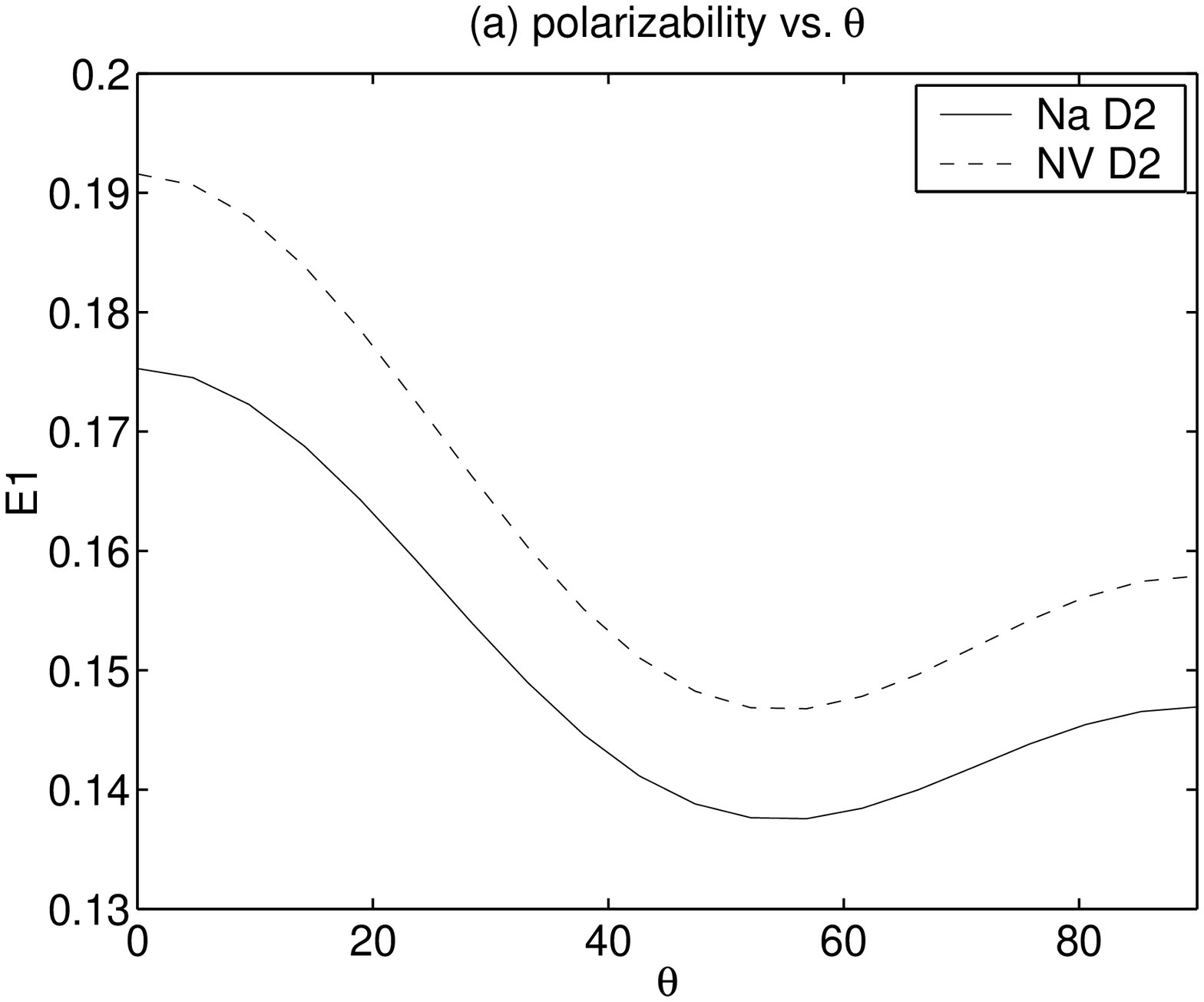}{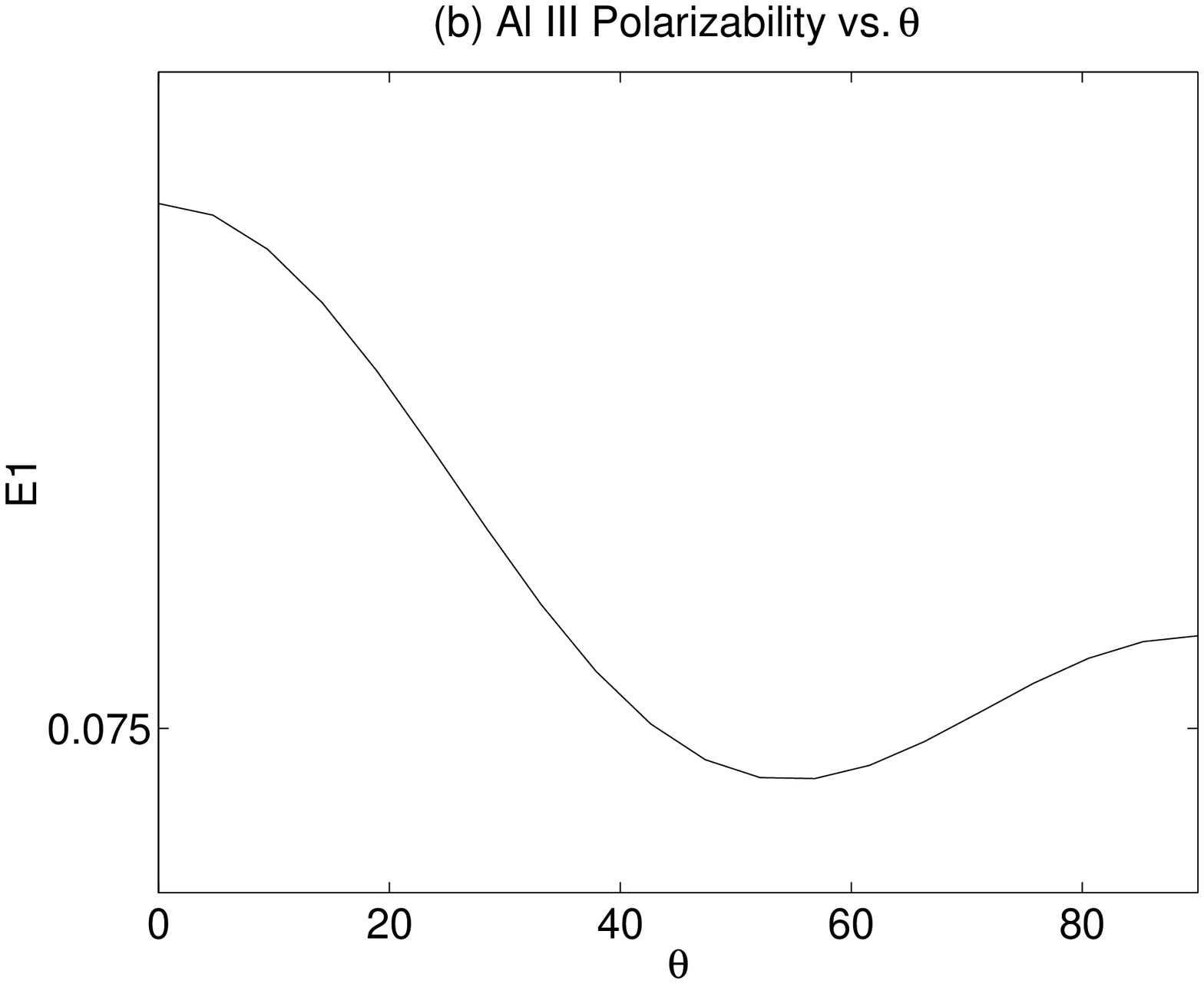}
\caption{(a) Polarizability E1 of Na D2 and NV D2 lines vs. the angle  $\theta$ between incident  
radiation and magnetic field; (b) Polarizability of Al III line vs. the angle  $\theta$.  
}
\end{figure}

{\bf AlIII}\\
Al III has a similar structure, but with a large nuclear spin $I=5/2$. Following the above procedure, 
we get the polarizability E1 for D2 line. It turns out to be very small ($<0.08$). Comparing E1 of all the sodium-like atoms, we can see that the more levels
the atom has, the less the polarizability is. This is understandable. As 
polarized radiation are mostly from those atoms with the largest axial angular momentum, which
constitute less percentage in atoms with more levels.

\begin{figure}
\label{Hsc}
\plottwo{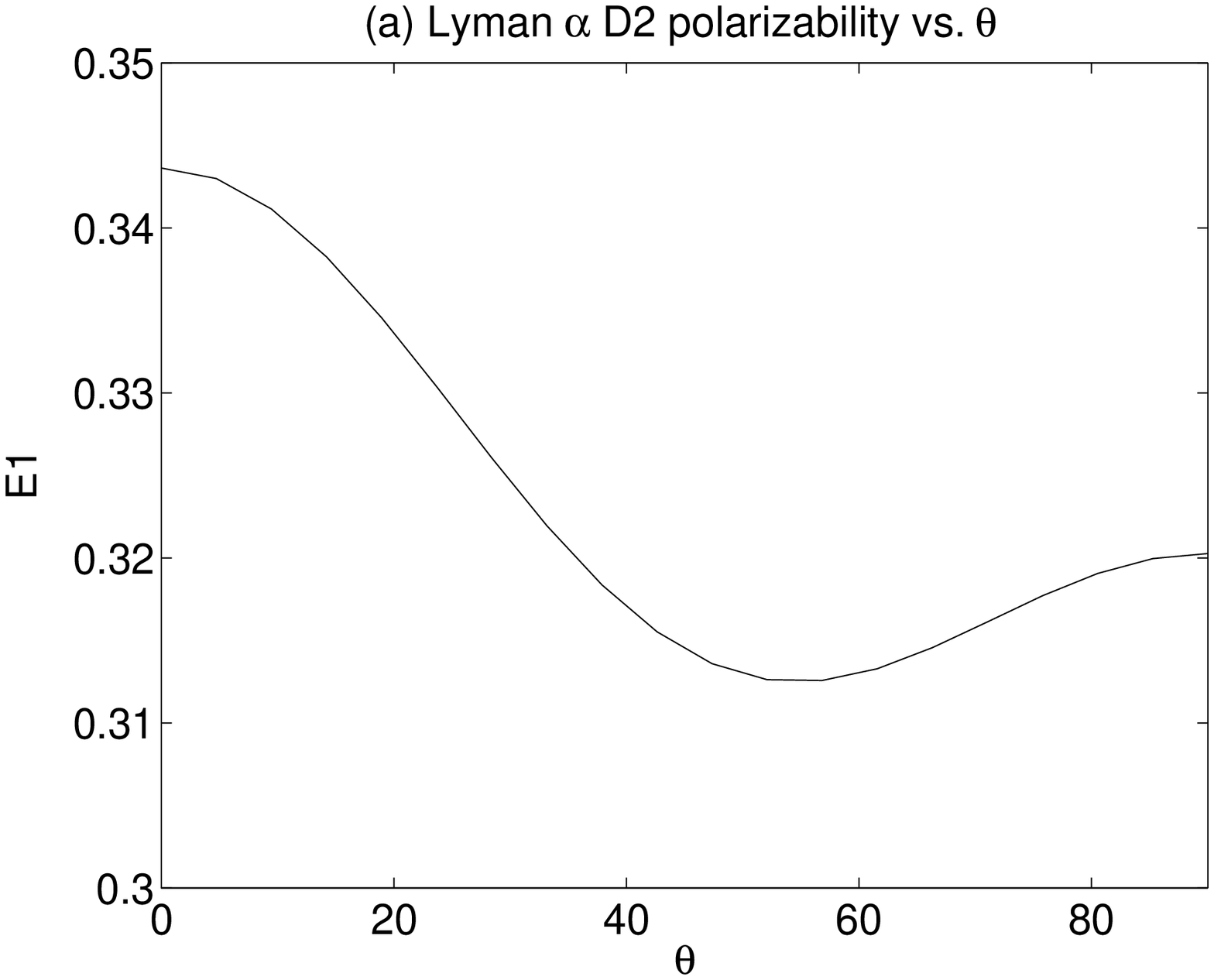}{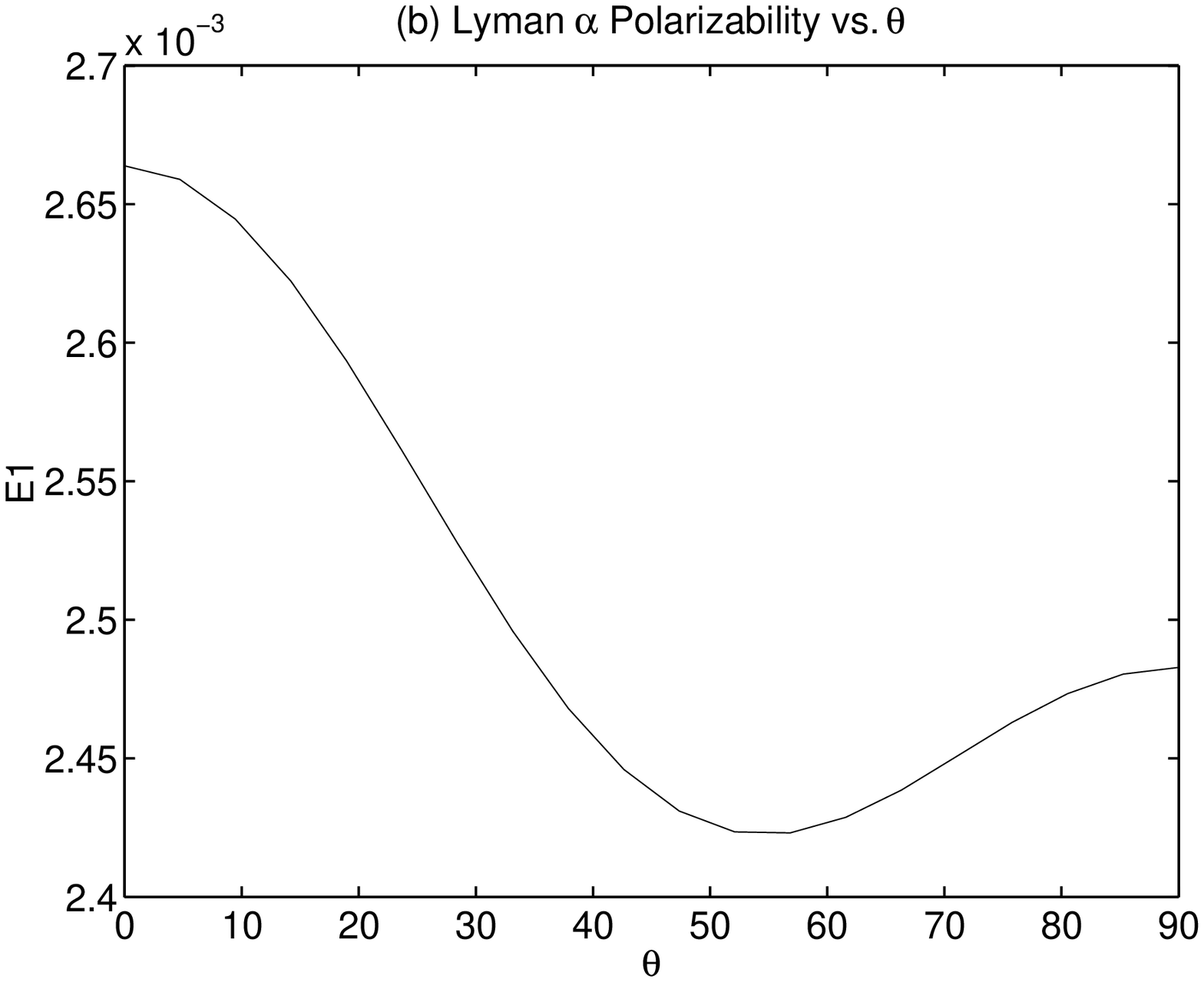}
\caption{Same as Fig.3a, but for Lyman $\alpha$: (a) D2 line; (b) unresolved (see text).
}
\end{figure}

{\bf HI}\\
More interesting example is Lyman $\alpha$ line of HI. The difference is that the 
hyperfine splitting of the excited state  is smaller than their natural energy width, e.g. 
$\omega_{nn'}=0.58\gamma$ for $2P_{1/2}$, and $\omega_{nn'}=0.23\gamma$ for $2P_{3/2}$. Thus 
the interference term in eq.(\ref{scatter}) must be taken into account. The nuclear spin of 
hydrogen is 1/2, thus $F_g=0, 1$ and there are totally four substates for ground state. 
The result for polarizability versus the angle between magnetic field and the incident light 
is shown in Fig.(\ref{Hsc}). Since the HI has the simplest structure comparing with other 
alkali atoms, its D2 line has the highest polarization according to the reason 
given in last paragraph (see Fig.\ref{Hsc}a). With present spectrometry, it's still difficult to resolve the D1 and D2 line especially because of Doppler shift. As a result of mixing with unpolarized D1 line, the polarizability is substantially reduced (see Fig.\ref{Hsc}b). 

\subsection{Alignment within fine structure}

Similar to species (atoms and ions) with hyperfine structure, the species with fine structure
can be aligned by radiation and realigned by magnetic field. The practical difference is that 
the transitions between different fine structure levels are much faster than those between
the hyperfine levels. If the decay among ground sublevels happens faster than the magnetic mixing, 
the distribution among different $J$ would be changing. However, the distributions among different 
$M_J$ will still only be affected by magnetic mixing.

{\bf NI and OII}\\
The ground state of O II is $4S^o_{3/2}$ and the excited state 
is $4P_{1/2,3/2,5/2}$. For the reason stated in $\S$3.2, the emission from $J=1/2$ is unpolarized. 
For the other two transitions, the results are shown in Fig.(\ref{O2ab}a). 

Neutral Nitrogen has the same structure as OII does, but with nuclear spin. 
Thus the transitions between hyperfine split levels should be considered in addition to those
between fine structure levels. As expected, this complex structure reduces the degree
of polarization in comparison with O II
\footnote{We did the calculation assuming the hyperfine splitting is at least three times larger 
than the natural linewidth and therefore the interference term is negligible for 
the excited state.} (see Fig.\ref{O2ab}b). 

The difference between them provides another possibility to measure magnetic field strength. 
When magnetic field is strong enough that Zeeman splitting becomes larger than hyperfine splitting, the electron and nuclear moments get decoupled. 
The alignment of NI then will follow the pattern of OII and the polarizabity becomes larger. 
Thus by the polarimetric study of NI, we can estimate magnetic field strength.    

\begin{figure}
\label{O2NI}
\plottwo{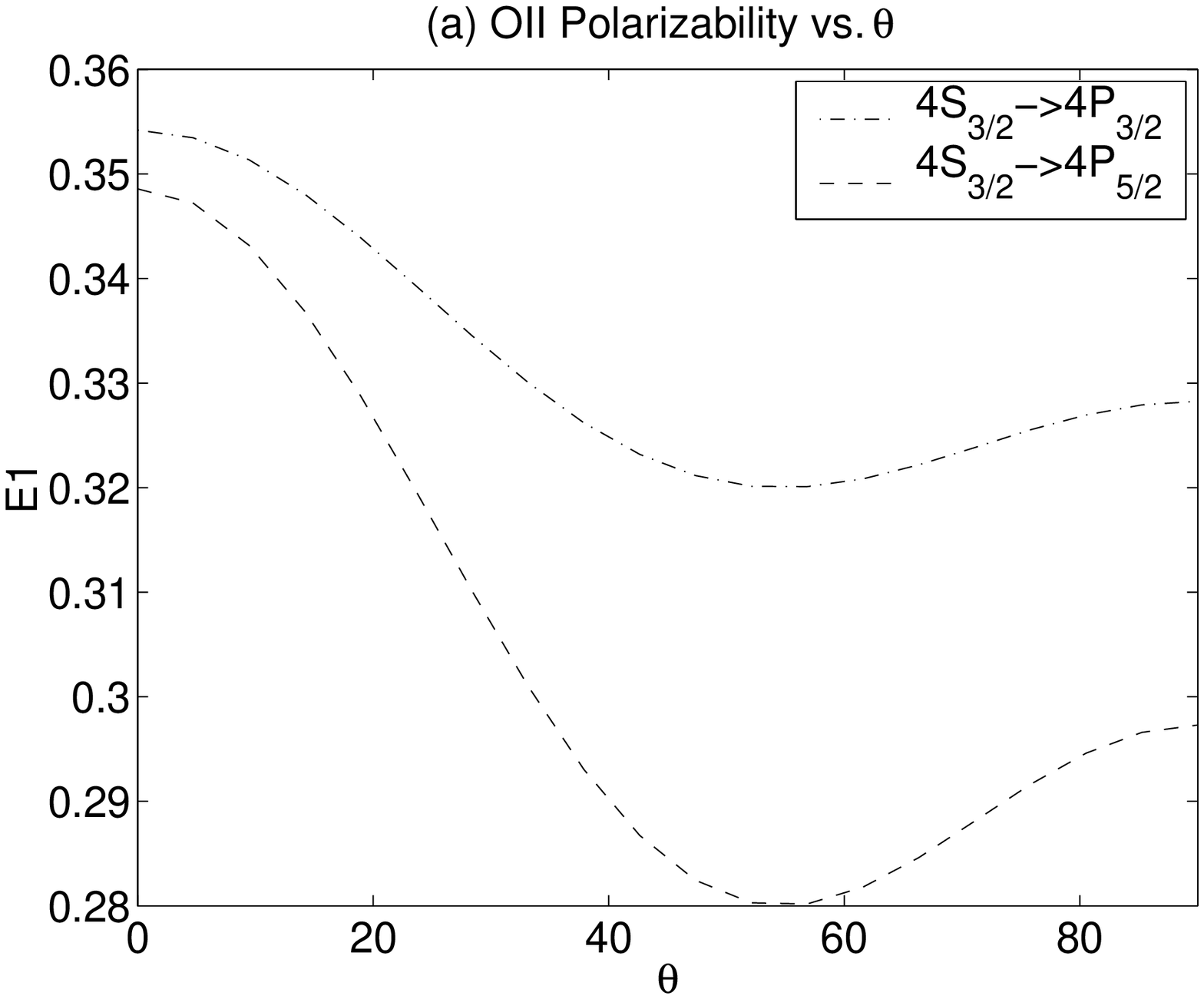}{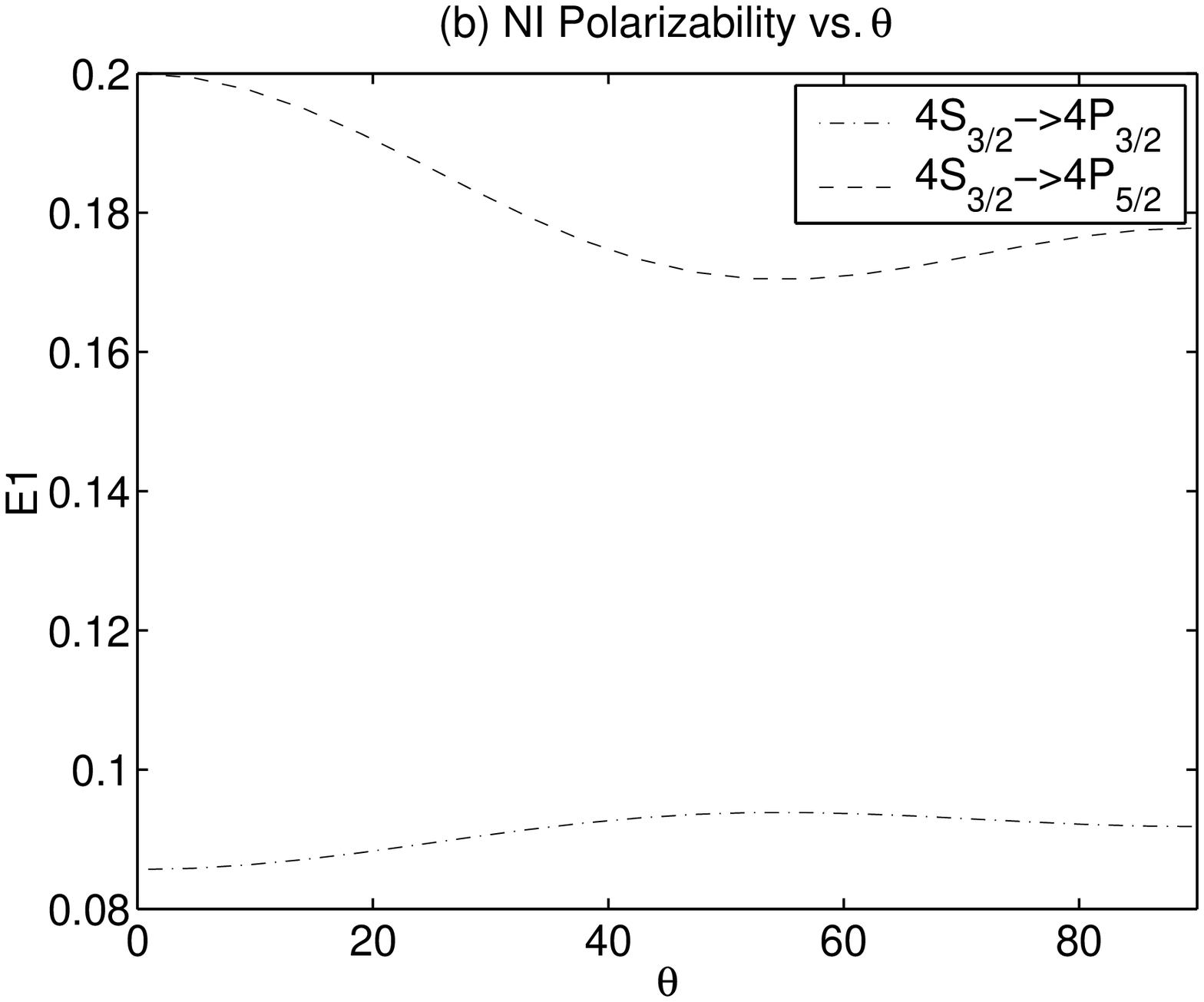}
\caption{Same as Fig.3a, but for (a) O II emission; (b) N I emission. The difference between them is due to the hyperfine splitting of NI (see text). 
}
\end{figure}

{\bf Cr II}\\
The ground term is $6S_{5/2}$, and the upper state is $6P^o_{3/2,5/2,7/2}$. This line is actually seen in QSO, so can be tested directly. The polarization of the two component $6S \rightarrow 6P_{3/2,5,2}$ varies marginally with the alignment. The result for the third component is shown in Fig.(\ref{mix1}a). More prominent effects are shown in absorption lines, which will be presented in \S4. 

\subsection{Effect of multiplets}
For multiplets, all the different transitions should be accounted according to their probabilities even if one is interested in one particular transition. This is because all transitions affect the ground populations and 
therefore the degree of alignment and polarization. 

It should be noted that even for one multiplet, all the transitions should be accounted within it in order to get the right degree of alignment and polarization%
\footnote{The only exception case is for those unpolarized lines, which have balance between all the three transitions $\Delta J=\pm1, 0$and alignment doesn't change anything}. The examples are given in \S3.1. Simply using the expressions  for the transitions 
from one J to another J (e.g. Stenflo's 1994, p188) will give wrong result. For instance, the ground state of O I is $3P_{0,1,2}$.  Permitted transitions can happen between it and $3S_1$ and $3D^o_{1,2,3}$. The transition probability for line $3P_{0,1,2}\rightarrow 3D^o_{1,2,3}$ is about ten time smaller than the other one, so we may safely ignore it. Nevertheless, all the transitions within the triplet of $3P_{0,1,2}\rightarrow 3S_1$ should be taken into account%
\footnote{ We assume equal distribution among these sublevels as the energy differences between these states are several hundred Kelvins}. The atoms can be excited from sublevels $3P_{0,1}$ and Raman scattered to $3P_2$ and change relative population among its magnetic sublevels and thus the polarization of scattered light from it. Because of the interplay among the triplet, the resulting polarization is marginal for the main component $3P_{0,1,2}\rightarrow 3S_1$.  However, if we use the equation in Stenflo (1994), we'll get a polarizability of 0.01.   

\begin{figure}
\label{mix1}
\plottwo{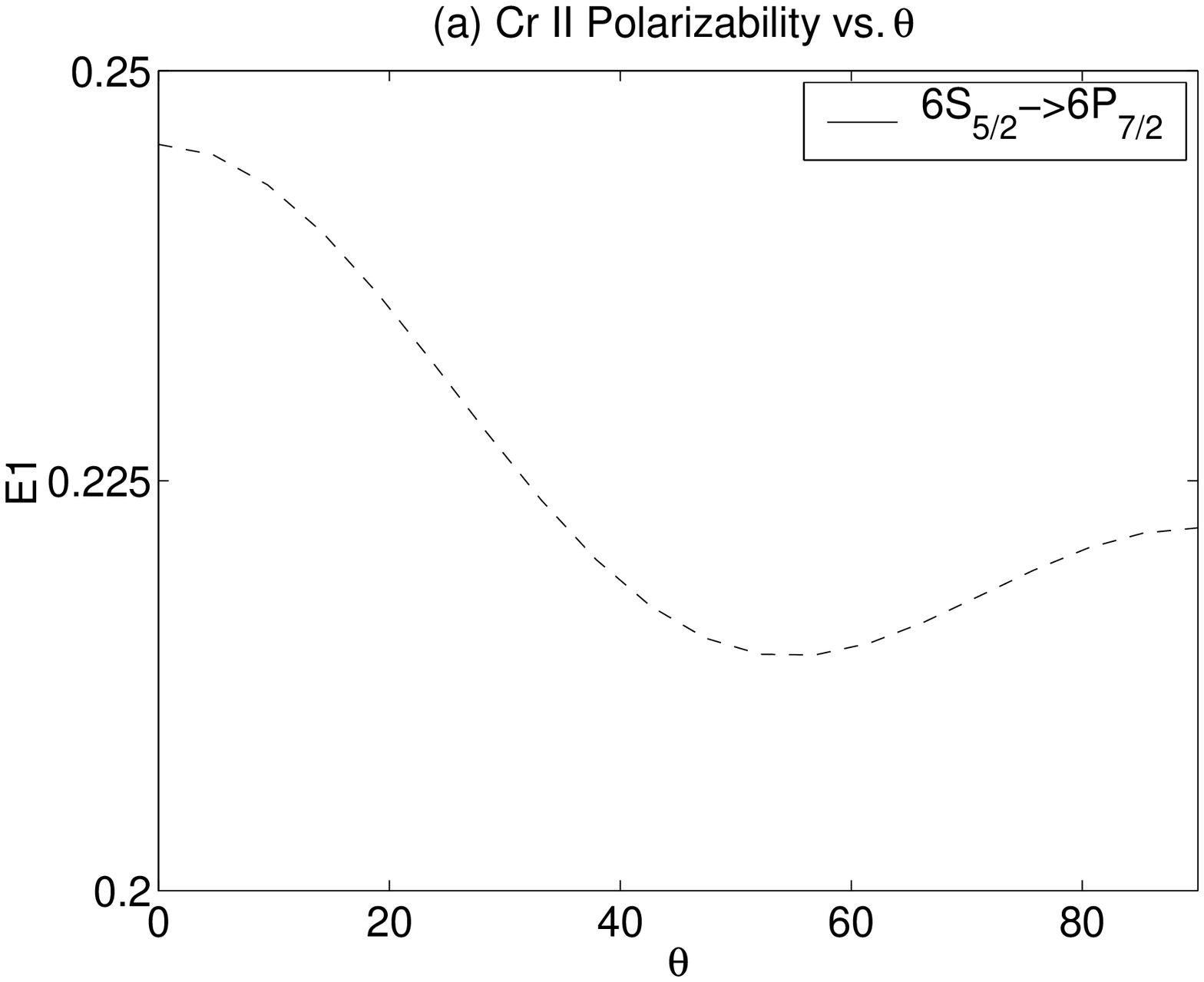}{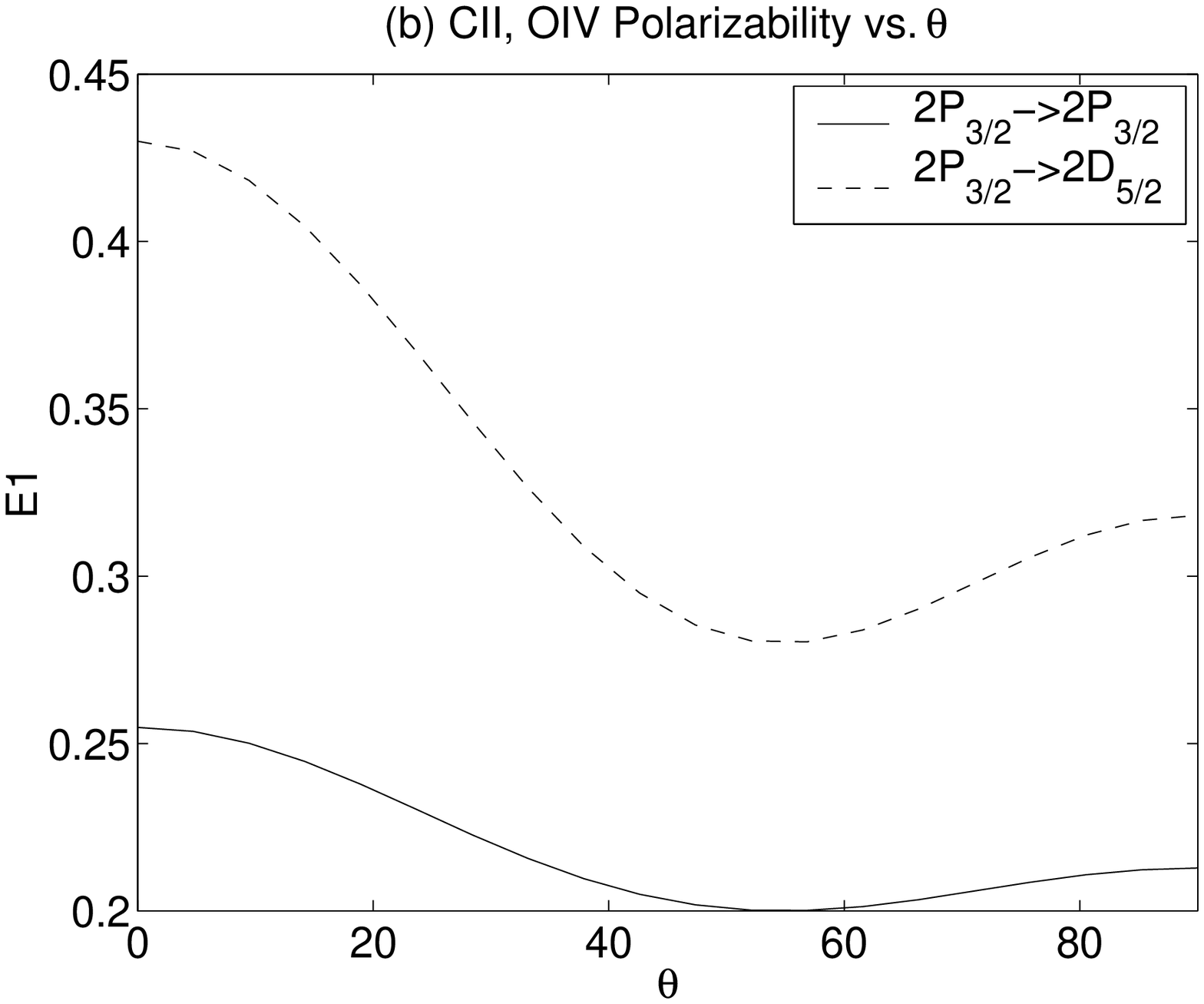}
\caption{Same as Fig.3a, but for (a) Cr II emission; (b) C II, OIV emission, the difference between CII and OIV is marginal as they have both similar structure and comparable relative transition probabilities between different multiplets (see text).}
\end{figure}

{\bf CII and OIV}\\
CII and OIV have the same structure. Their ground states are $2P^o_{1/2,3/2}$ and excited 
states are $2S_{1/2}, 2P_{1/2,3/2}, 2D_{3/2,5/2}$. For multiplets, both of the atoms, the transition from 
$2P^o\rightarrow 2D$ transition is dominant, so their degrees of polarization are nearly the same (see
fig.\ref{mix1}b).

\begin{figure}
\label{CIO3}
\plottwo{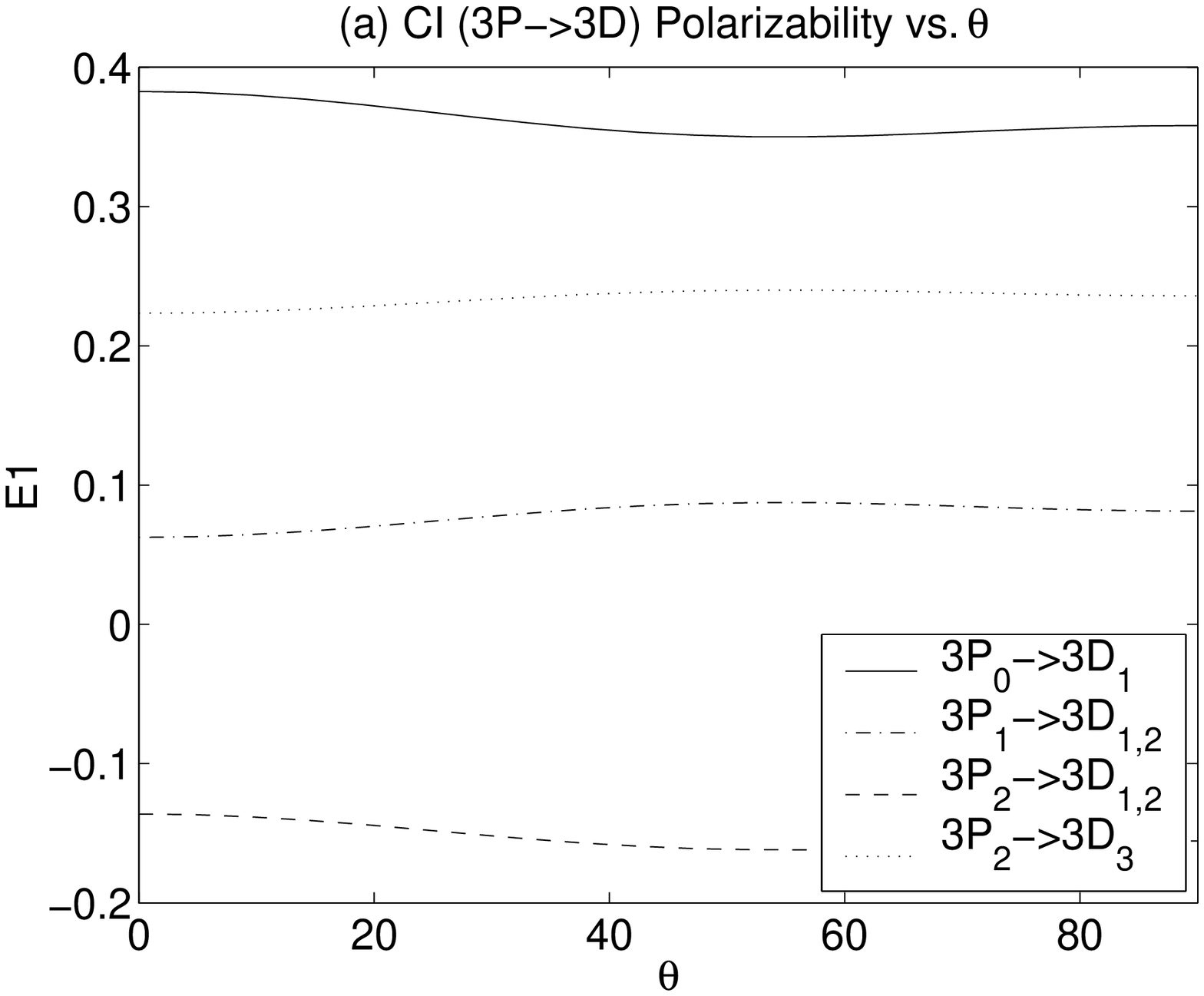}{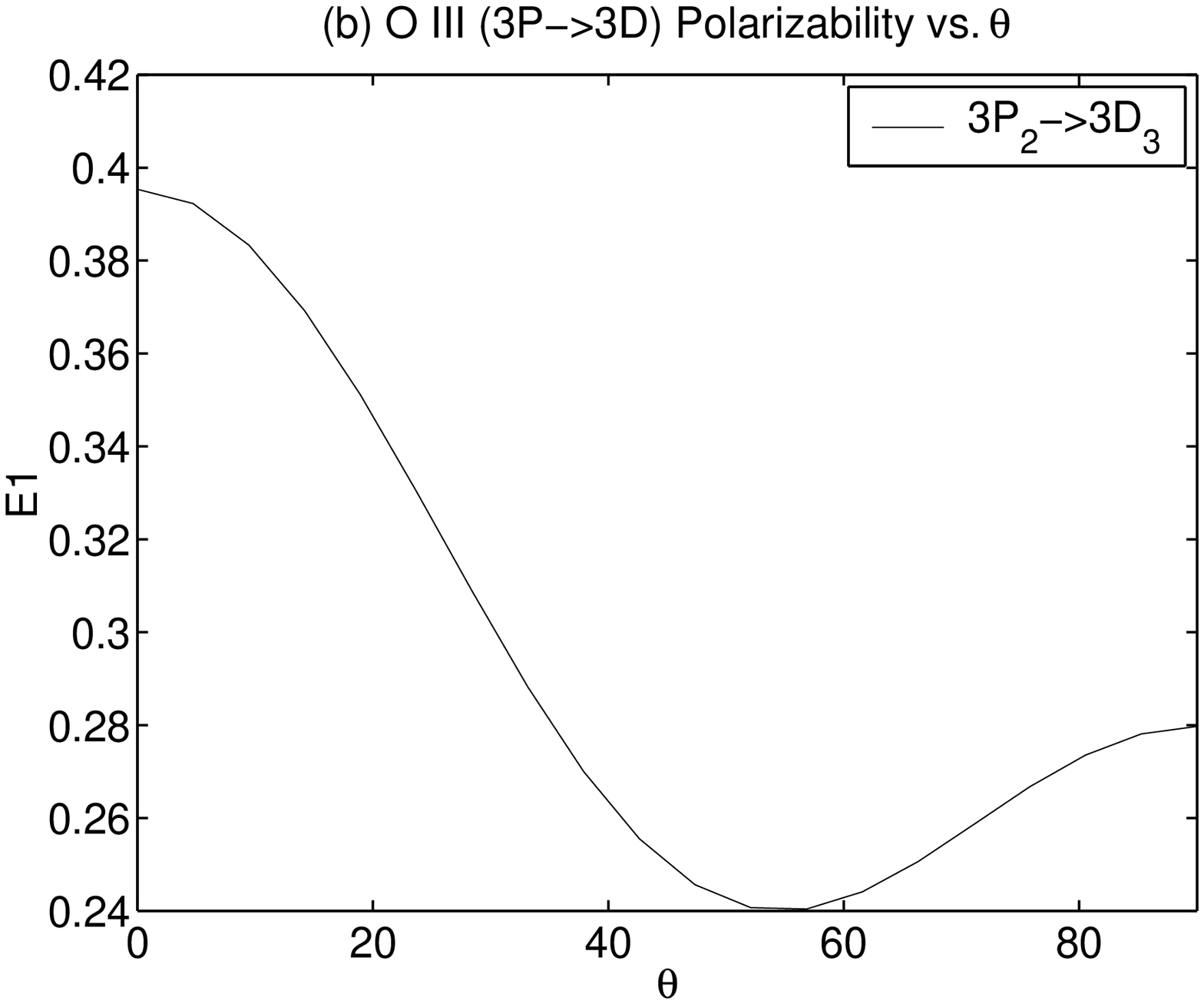}
\caption{Same as Fig.3a, but for (a) C I emission, the scale range is larger in this plot as the polarizability of the four lines have very different amplitudes, as a result the variations are not so obvious although they are comparable with those for other atoms; (b) O III emission. The difference between CI and O III is merely due to multiplet effect (see text).
}
\end{figure}

{\bf CI and OIII}\\
The situation with CI and OIII is different. They have similar structure. 
Their difference is primarily due to multiplet effect. For OIII, the probabilities of 
transitions $3P\rightarrow 3P^o$ and $3P\rightarrow 3D^o$ are comparable, while C I is 
dominated by $3P\rightarrow 3P^o$. Thus their ground populations differ after the pumping by incident radiation. As a result they have different polarizations (see  Fig.\ref{CIO3}).

\begin{table*}
\begin{tabular}{||c|c|c|c|c|c|c||}

\hline
 Atom&
 Nuclear spin&
 Lower state&
 Upper state&
 Wavl(\AA)&
 Pol(emi)&
 Pol(abs)\tabularnewline
\hline 
\hline 
{Na I}&
$\frac{3}{2}$&
$1S_{1/2}$&
$2P_{3/2}$&
5891.6&
Y&
Y (if resolvable)\\
\cline{4-7}
& &  &
$2P_{1/2}$&
5897.6&
N&
Y (if resolvable)\\
 \hline
{N V}&
$1$&
  $1S_{1/2}$&
$2P_{3/2}$&
1238.8&
Y&Y (if resolvable)\\
\cline{4-7}
& & &
$2P_{1/2}$&1242.8&N&Y (if resolvable)\\
\hline
 Al III&
$\frac{5}{2}$&
$1S_{1/2}$&
$2P_{3/2}$&
 1854.7&M&M\\
\cline{4-7}
&& &
$2P_{1/2}$ & 1862.7&N&M\\
\hline
 H I&
$\frac{1}{2}$&
$1S_{1/2}$&
$2P_{1/2,3/2}$&
912-1216&Y&N\\
\hline
 N I&
1&
$4S^o_{3/2}$&
$4P_{1/2,3/2,5/2}$&
$865-1201$&Y&Y\\
\cline{1-2} \cline{5-5}
 O II&
0& & &
 375-834& & \\
\hline
 O I&
0& 
$3P_2$&
$3S_1$&
911-1302.2&M&Y\\
\hline
Cr II&0 &$6S_{5/2}$ & 
$6P^o_{3/2,5/2,7/2}$&
2056, 2062, 2066&Y
&Y \\
 C II&
0&
$2P^o_{3/2}$&
$2P_{3/2}$&
904.1&Y&Y\\
\cline{4-5}
& & &
$2D_{3/2,5/2}$&
1335.7& & \\
\hline
O IV&
0&
$2P^o_{3/2}$&
$2P_{3/2}$&
554.5&Y&Y\\
\cline{4-5}
& & &
$2D_{3/2,5/2}$&
239, 790& & \\
\hline
 C I&
0&
$3P_{1,2}$&
$3P^o_{0,1,2}$&
1118-1657&Y&Y\tabularnewline
\cline{4-5}
 &
 &
 &
$3D^o_{1,2,3}$&
 1115-1561& & \\
\hline
O III&0 &$3P_{0,1,2}$ & 
$3P^o_{0,1,2}$&
304, 374, 703&Y
&Y \\
\cline{4-5}
& & & $3D^o_{1,2,3}$&267, 306, 834& &\tabularnewline
 \hline
\end{tabular}
\caption{Note: only polarizable and alignable components are listed. "M" stands for marginal polarization. Those lines with polarizability or polarization or their variations less than 1\% are not plotted in this paper.}
\end{table*}

\begin{figure}
\label{Naab}
\plottwo{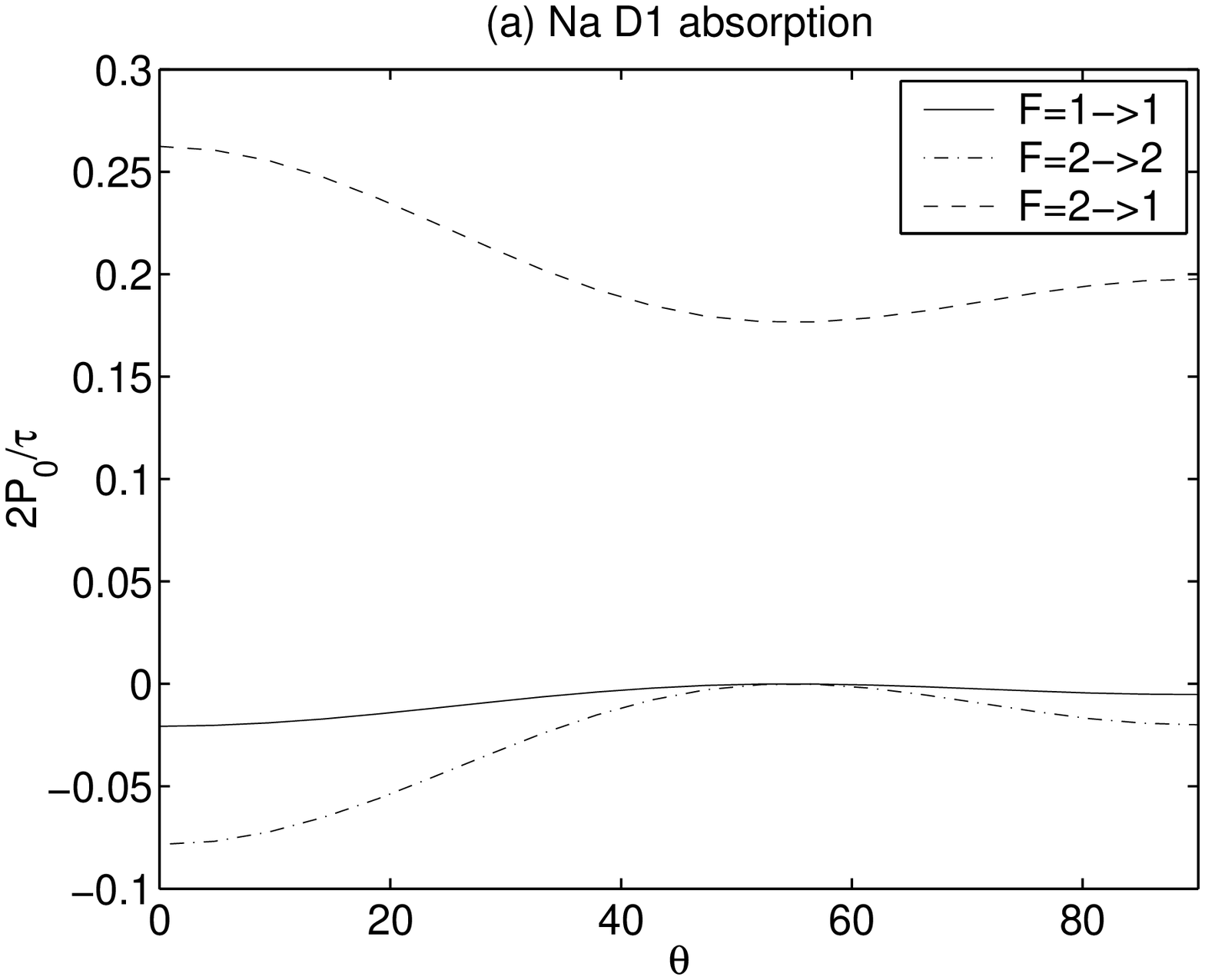}{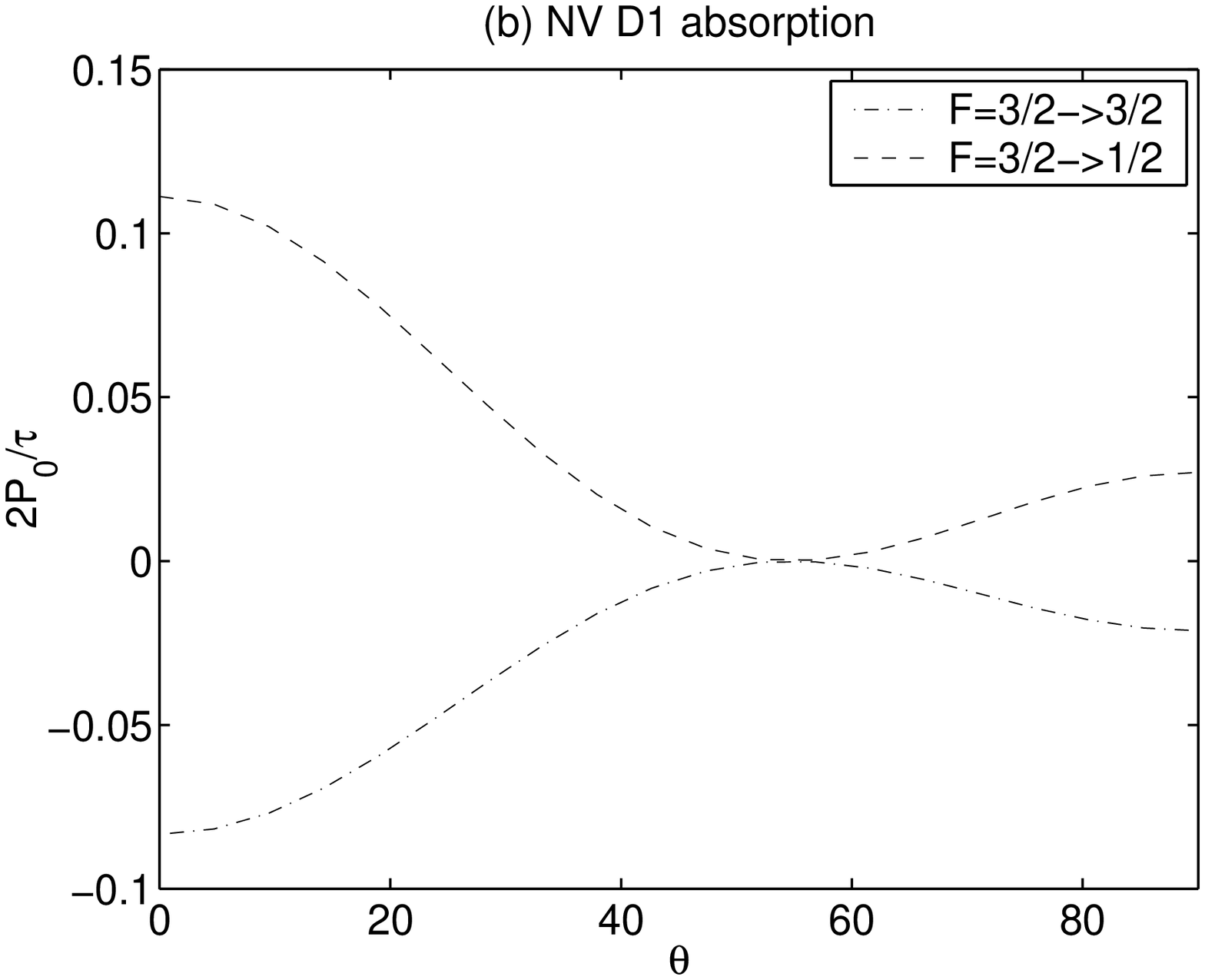}
\caption{The ratio of polarization to optical depth versus the angle $\theta$ between local magnetic field and the pumping light for (a) Na I absorption; (b) NV absorption.} 
\end{figure}

\begin{figure}
\label{O2ab}
\plottwo{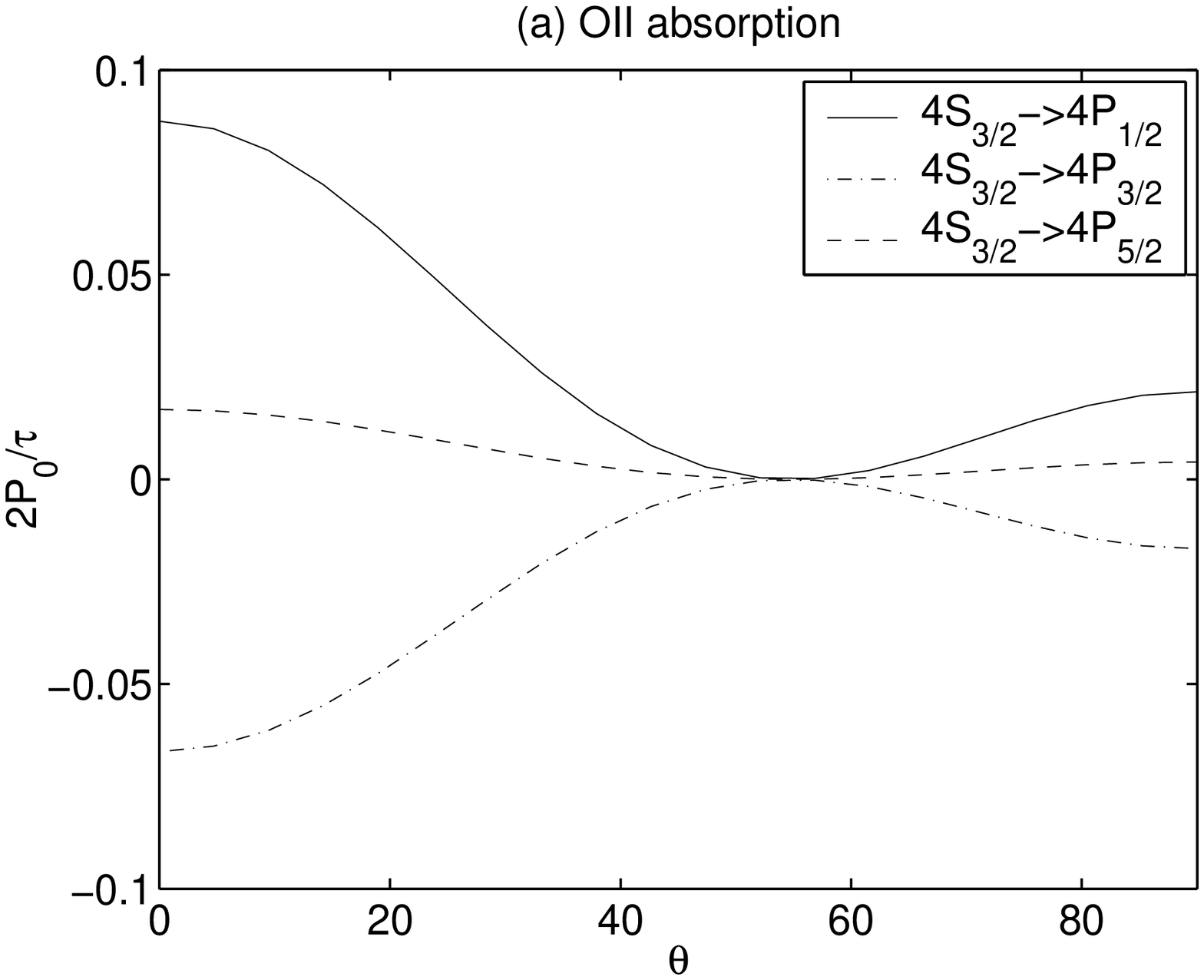}{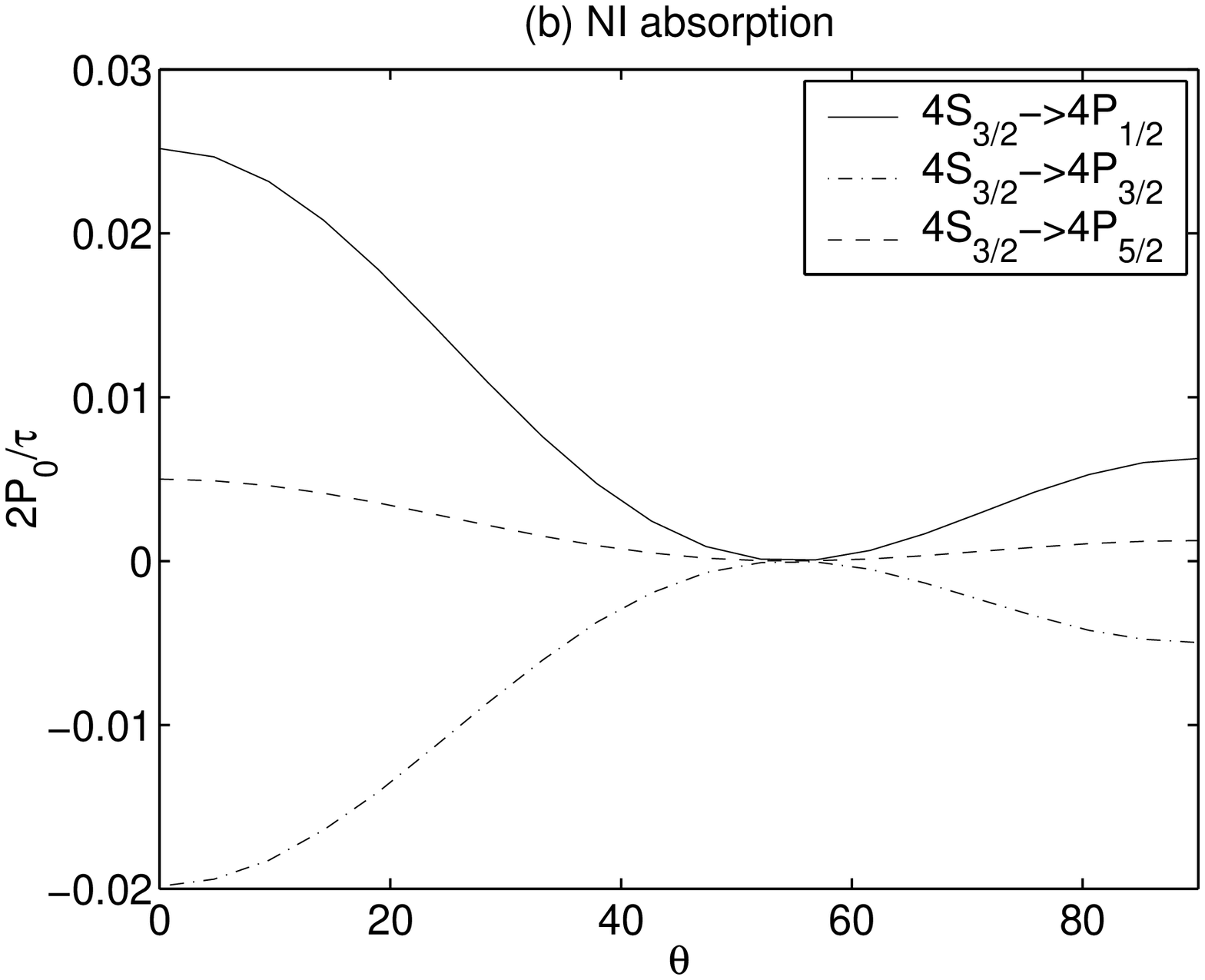}
\caption{Same as Fig.7, but for (a) O II; (b) NI, the difference between them arises from the fact that NI has hyperfine splitting.} 
\end{figure}

\begin{figure}
\label{CIab}
\plottwo{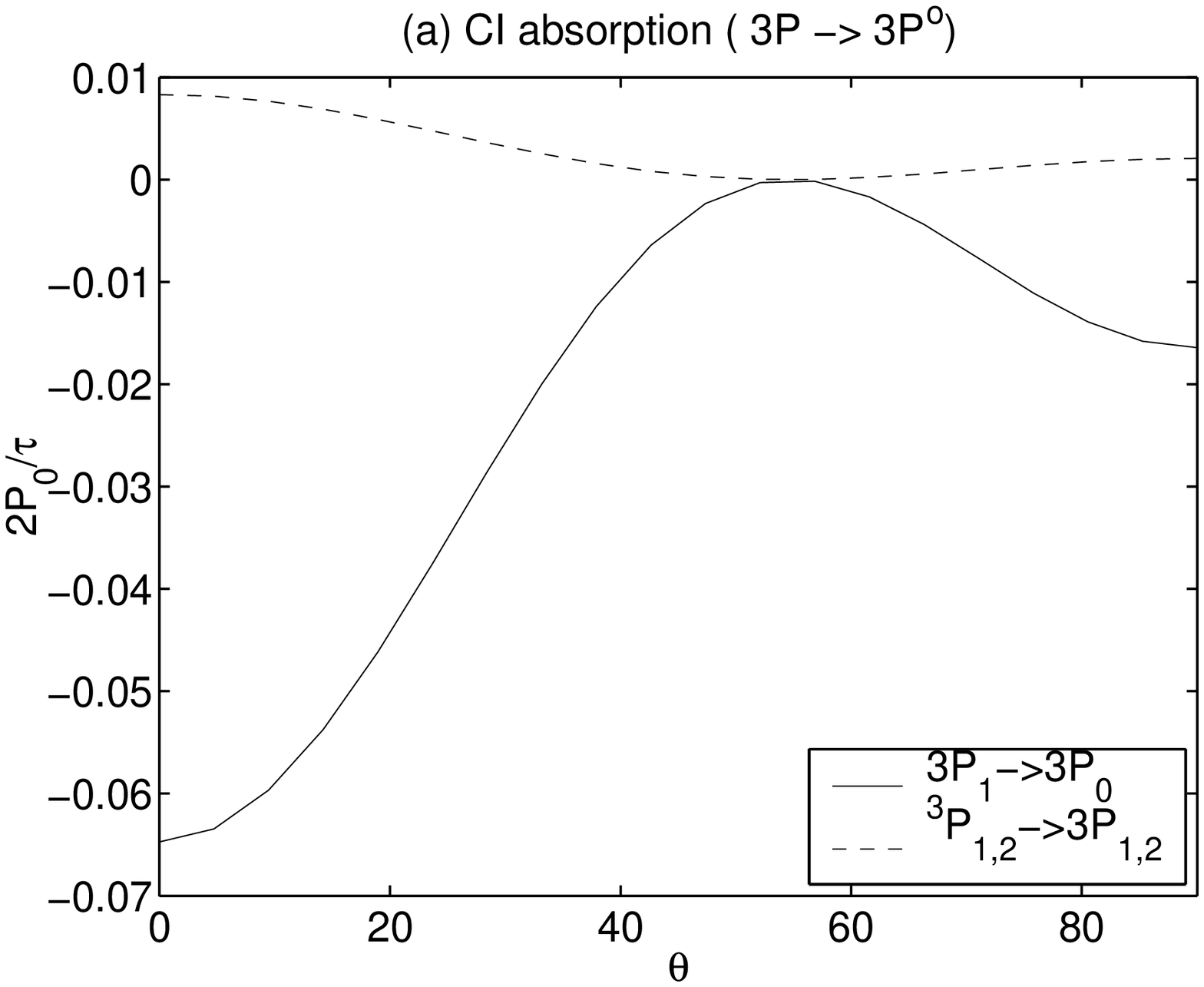}{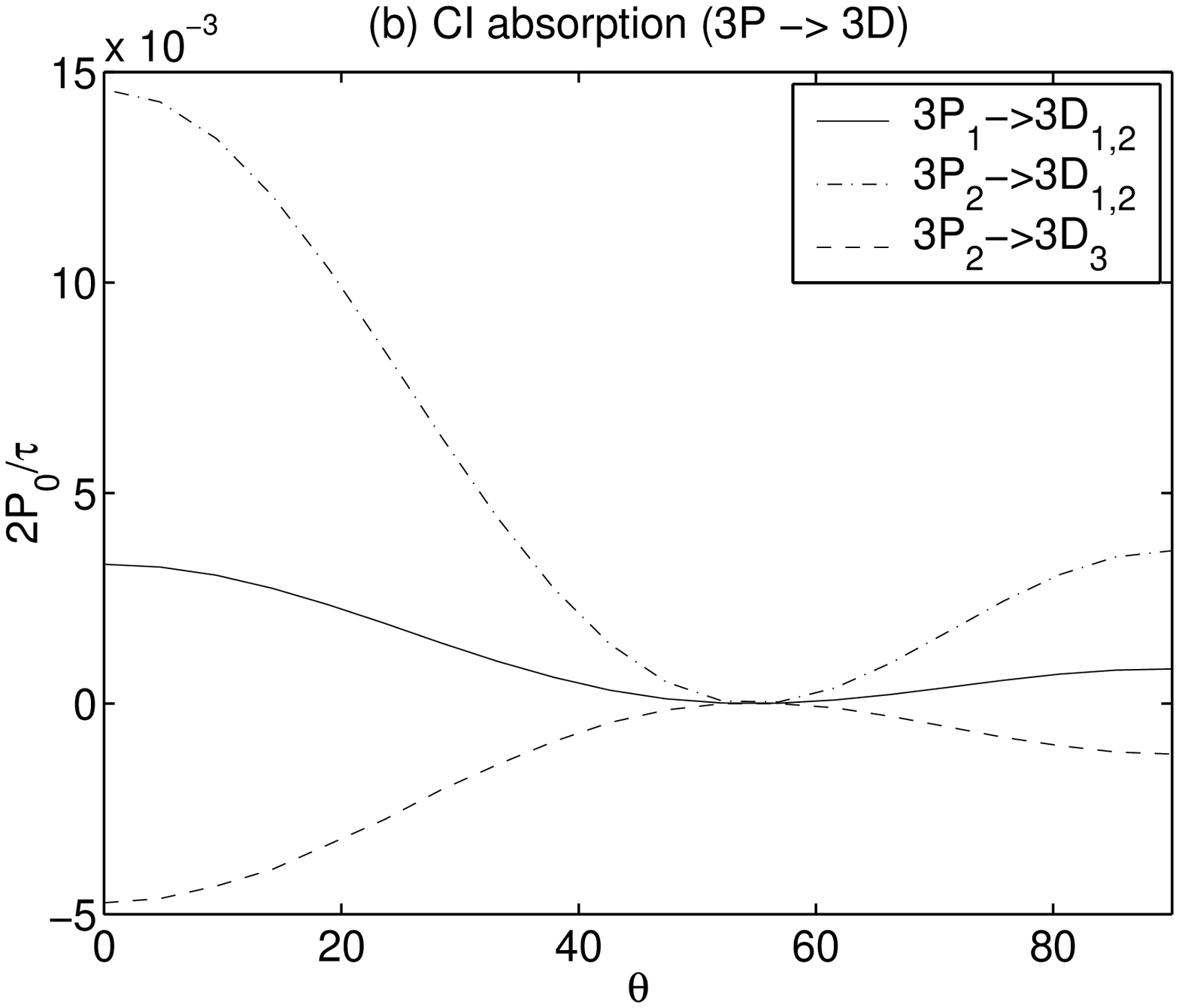}
\caption{Same as Fig.7, but for (a) C I (3S-3P); (b) CI (3S-3D).} 
\end{figure}

\begin{figure}
\label{O3ab}
\plottwo{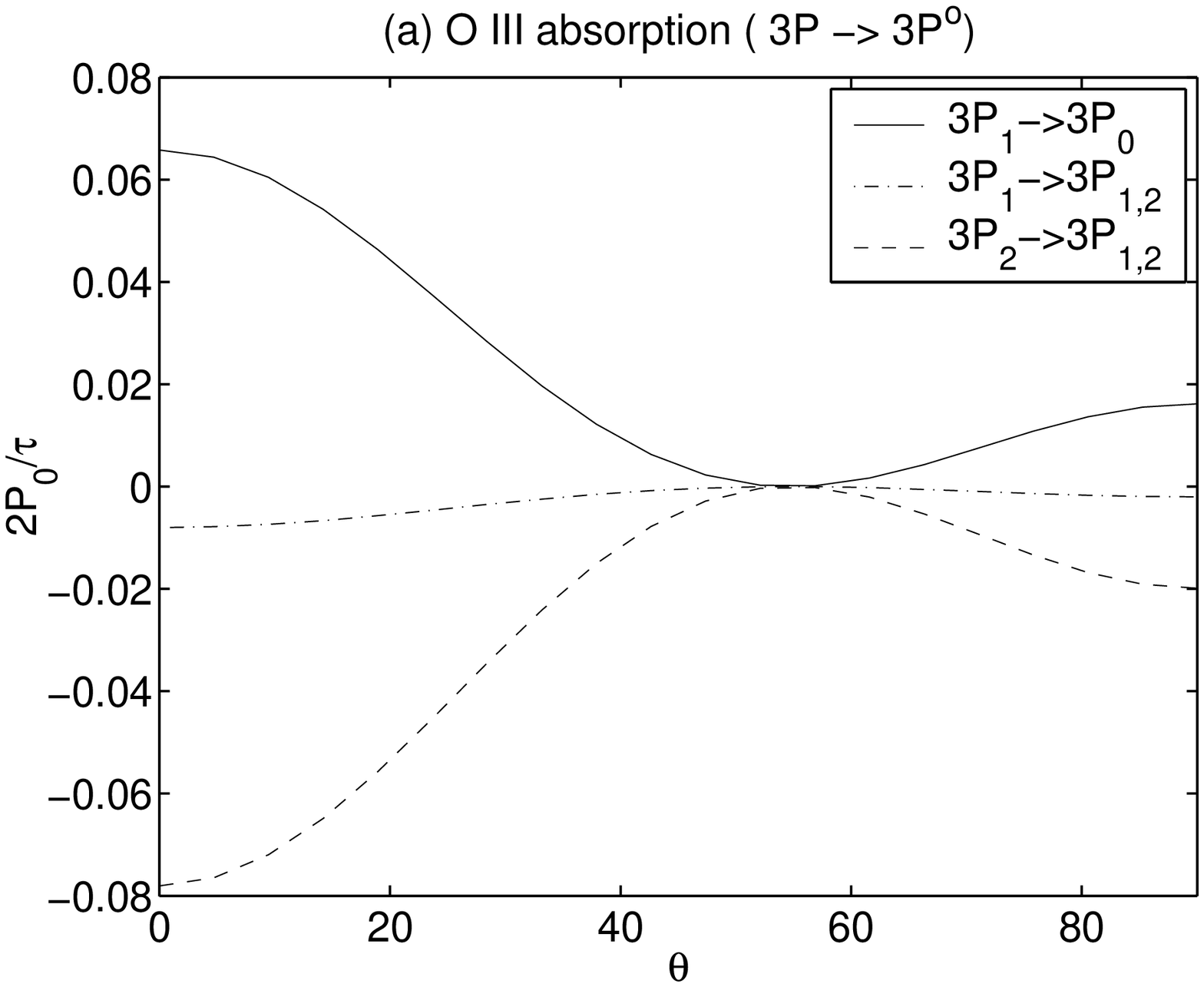}{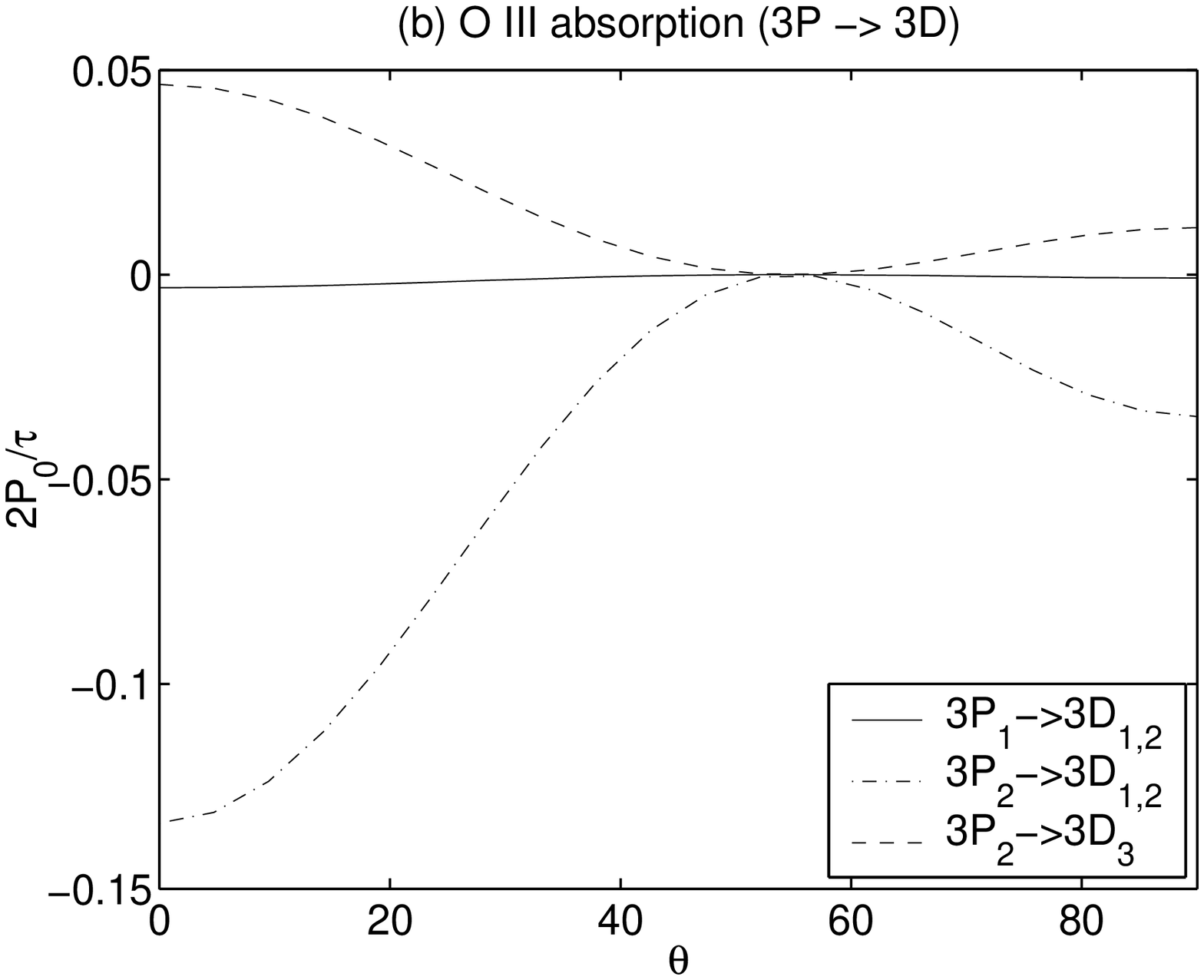}
\caption{Same as Fig.7, but for (a) O III (3S-3P); (b)  OIII (3S-3D).} 
\end{figure}

\begin{figure}
\label{mix2}
\plottwo{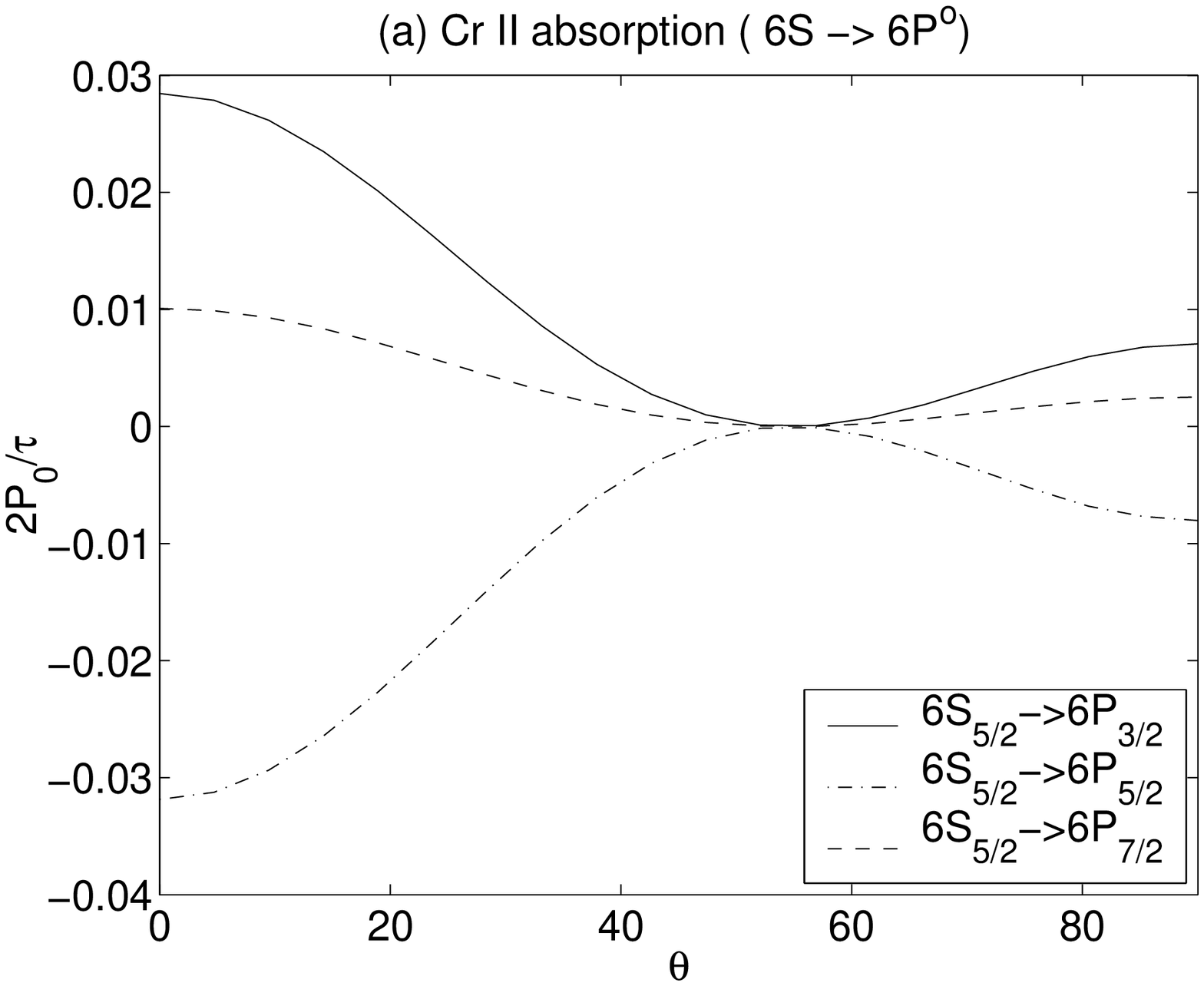}{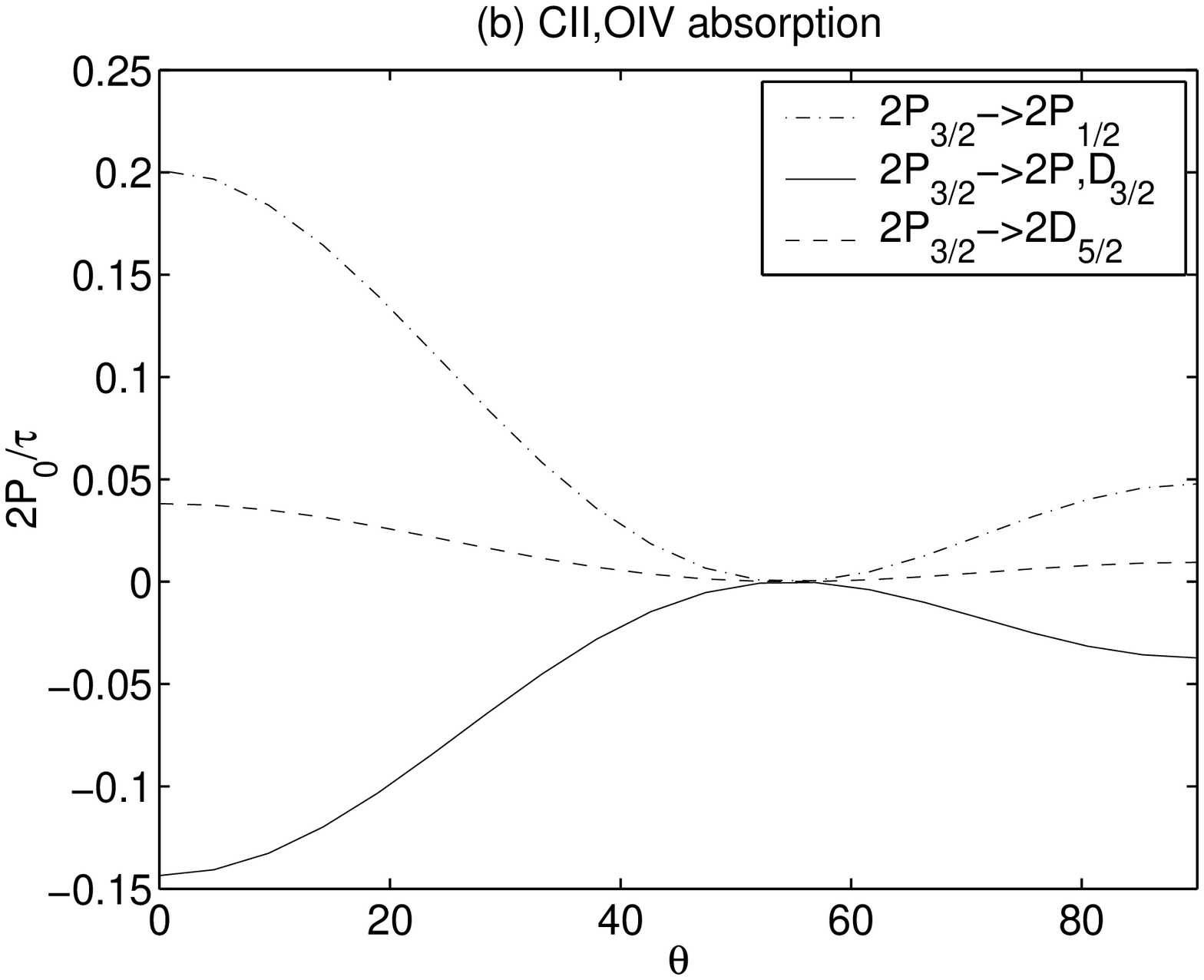}
\caption{Same as Fig.7, but for (a) Cr II; (b) CII, OIV, similar to emission case, the difference between CII and OIV is marginal as they have both similar structure and comparable relative transition probabilities between different multiplets.
} 
\end{figure}

\subsection{Effect of Multiple Atom-Photon Interactions: Saturated State}

\begin{figure}
\label{satura}
\plottwo{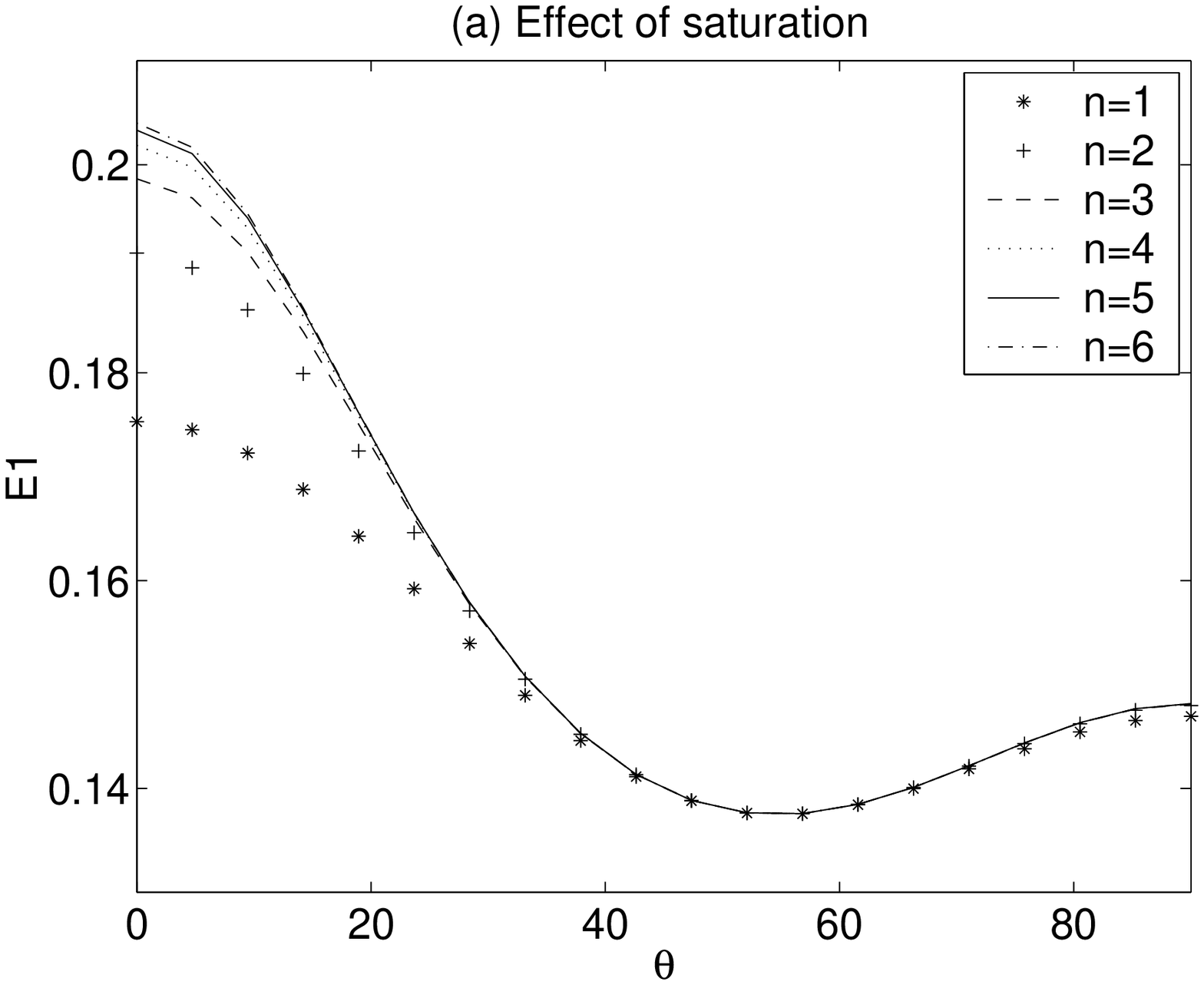}{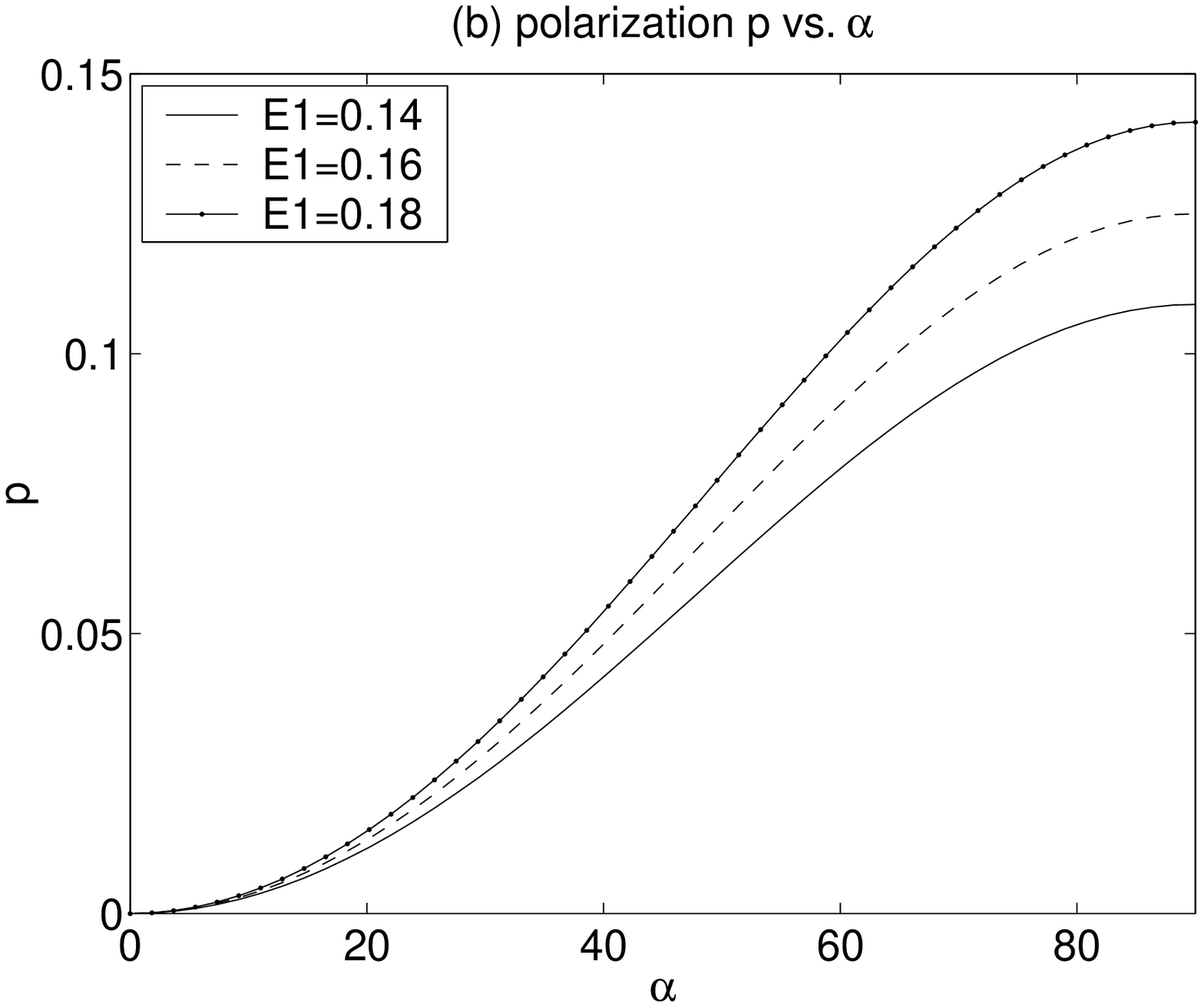}
\caption{(a) The polarizability of Na D2 line from atoms pumped 'n' times; (b) Polarization vs. the scattering angle $\alpha$ 
(see eq.\ref{polarization})}
\end{figure}

The atomic alignment and the degrees of polarization that it entails have been calculated so
far assuming that atoms resonantly scatter one photon. 
Atoms can scatter a few photons, i.e. be pumped multiple times, and this changes the degree of alignment.
In this case, the occupation vector should be multiplied first by $S_{gg'}$ and then by $B_{gg'}$. 
The resulting alignment and the degree of polarization that it entails will increase 
with the number of the scattering events. The effect  of many photon interactions gets to saturation,
however. Using sodium D2 line as an example, we calculate the polarizations for different 
numbers of scattering events (see Fig.\ref{satura}). Apparently the less the mixing is, 
the faster the degree of alignment will increase with the number of scattering events. 
We see also from Fig.(\ref{satura}) that the alignment gets quickly saturated, 
after approximately four scattering events 
more scattering doesn't cause more alignment and more polarization. 
This means for a sufficiently high rates of photon arrival, the polarization has 
 one to one correspondence with the angle 
$\theta$. 

\section{Polarization of Absorbed Radiation}

So far we have discussed the situation of polarization of the scattered radiation.
Light absorbed by aligned atoms will also be polarized as well. In practical terms
this means that there is a strong source of radiation that aligns atoms. The light
of a weak source can be polarized while passing through the cloud of aligned atoms.
Absorption by the strong source that aligns atoms is also affected by the alignment.

 We can obtain the absorption coefficients using Eq.(\ref{decomp})
\begin{eqnarray}
\mu_{\parallel}&=&\Sigma_{g} \rho_g ((R_{ge}^1+R_{ge}^{-1})^2\cos^2\alpha/2+(R_{ge}^0\sin\alpha)^2),\nonumber\\
\mu_{\perp}&=&\Sigma_{g} \rho_g ((R_{ge}^1+R_{ge}^{-1})^2/2.
\label{mupp}
\end{eqnarray}

In optical thin case, the absorbed radiation has a polarization of

\be
P_{abs}=\frac{e^{-\tau_\parallel}-e^{-\tau_{\perp}}}{e^{-\tau_\parallel}+e^{-\tau_{\perp}}}\simeq (\tau_\parallel-\tau_\perp)/2=\frac{\tau(1-r)}{2(1+r)},
\label{Pabs}
\ee
where $\tau=\sqrt{\mu_\parallel^2+\mu_\perp^2}$, $r=\mu_\parallel/\mu_\perp$. According to this, the polarization of light absorbed 
at arbitrary angle $\alpha$ is related to the polarization at $\alpha=90^o$ by

\be
P_{abs}=\frac{P_{abs}(90^o)\sin^2\alpha}{1+P_{abs}(90^o)\cos^2\alpha},
\ee 
where $P_{abs}(90^o)$ is obtained by combining Eqs.(\ref{mupp}) and (\ref{Pabs}). Since usually the excited level has more substates than the ground level, 
there are more routes for absorption. As a result, the chances to absorb different 
polarized photons is more likely to be equal. {\it Therefore, different components with different 
upper sublevels need to be resolved to get the polarization}. While this may be challenging for 
hyperfine components with present technique (see Figs.\ref{Naab}, \ref{CIab}a ), there shouldn't 
be a problem with fine components (see Figs.\ref{O2ab},\ref{CIab}b,\ref{O3ab} and \ref{mix2}).

\section{Earlier work on atomic alignment}
The work on neutral sodium laboratory alignment was pioneered more that half a 
century ago by Brossel, Kastler and Winter (1952) and H54. These experiments revealed that sodium atoms
can be efficiently aligned in laboratory conditions if atomic beams are subjected to anisotropic 
resonance radiation flux. 

General discussion of the atomic
alignment, in particular, of HI can be found in the works of Varshalovich (1967). In Varshalovich (1967) the variations of the absorption by HI due
to atomic alignment are
considered. However, the effects of magnetic realignment were not discussed
in these works.

Landolfi \& Landi Degl'Innocenti (1986) for the first time investigated the possibility of using atomic alignment to detect weak magnetic field in diffuse media. They called the effect "lower level Hanle effect" as comparing with Hanle effect on upper levels. They modeled fine structure alignment of two level atoms. In terms of magnetic field, they considered only two cases, along line of sight or away from it.  

Lee \& Blandford (1997) discussed alignment of atoms with fine structure
in relation to quasars. Again the effects of magnetic field was not
taken into account and therefore an interesting possibility of
studying magnetic field of quasars through polarimetry was not considered.

It happened rather recently that the effect
has been rediscovered by Solar researchers (see Trujillo Bueno \& Landi Degl'Innocenti
1997). Polarimetry of the line emission have been proven to be so
informative and a term \char`\"{}second solar spectrum\char`\"{}
has been coined by Stenflo Keller (1997) to reflect the importance
of the new window for Solar studies. 
The studies of the polarization resulted in important
change of the views on Solar corona (Landi Degl'Innocenti 1998, Trujillo
Bueno 1999, 2002, Trujillo Bueno et al. 2002, Casini et al. 2002, Manso Sainz \& Trujillo Bueno 2003). However, these studies correspond to a setting different from
what we consider here.
In this paper we are concentrated on weak field regime, in which is the ground level populations of atoms that are influenced by magnetic field and determine the polarization. This is the case for interstellar medium. The polarization of absorption lines that we found is thus more informative. 

\section{Discussion}

We have considered the situation that atoms are subjected to the 
flow of photon that excite transitions at the rate $\tau^{-1}_A$ which is
smaller than the Larmor precession rate $\tau^{-1}_L$, but larger or comparable
with the rate of disalignment due to collisions $\tau^{-1}_R$. For the cold
gas with 10 atoms per cubic cm, temperature of 100 K and magnetic field
of $6\times 10^{-6}$G the characteristic range over which the atoms can be
aligned by an O star is $3.2\times 10^{3}\sim 2.7\times 10^8 R_\odot$ (or 20 Au) for NaI.

A detailed discussion of alignment of atoms in different conditions 
corresponding to 
circumstellar regions, AGNs, Lyman alpha clouds, interplanetary space
is given in our subsequent paper.

Note, that when the time of the photon arrival gets of the order of
Larmor frequency the polarization gets sensitive to the amplitude of
magnetic field. This opens an avenue of obtaining the {\it strength}
of weak magnetic fields. The complication is that this effect may not be 
easily 
disentangled with magnetic mixing. If there are several species spatially 
correlated, 
however, it's possible to measure both the magnetic field strength and
 direction. 

Both absorption and scattering can be used to study astrophysical
magnetic fields. Scattering when the radiation source is localized 
and its relative 
position is well known,
as this is the case of a sodium tail of a comet, presents the easiest case for 
polarimetric study. The geometry is also known for most for scattering 
studies of circumstellar regions. 

For absorption studies, there can also be situations where the light
passes near a radiation source that creates alignment. 
In general, for absorption studies
the direction to the source of aligning radiation is another variable.
We discuss in a subsequent paper,
this variable can be defined either using different lines.
Atomic alignment can provide unique information about magnetic field.
For instance, grain alignment\footnote{Grain have tendency to be aligned 
with their long
axes perpendicular to magnetic field.} (see review by Lazarian
2003) can provide information about the component of magnetic field
perpendicular to the line of sight. The same is true in relation to
the Goldreich-Kylafis effect\footnote{Atomic alignment has some similarity to
 Goldreich-Kylafis effect, which also measures magnetic field through 
magnetic mixing. 
However, Goldreich-Kylafis effect is concerning the polarization of radio 
lines. 
The upper states of radio lines are so long lived that significant magnetic 
mixing can happen 
among  different magnetic sublevels of these states. 
Atomic alignment, on the other hand, happens with ground states 
of optical and UV transitions.} (Goldreich \& Kylafis 1982, Girart 1999).
Atomic alignment provides the component of magnetic field (angle $\theta$) 
that is not
measurable otherwise, which opens new perspectives for tomography
of 3D magnetic fields.

Note, that for interplanetary studies,
one can investigate not only spatial, but also temporal variations
of magnetic fields. This can allow cost effective way of studying
interplanetary magnetic turbulence at different scales.

To finish, let us mention a fact not related to the major
thrust of our paper, which is a development of a tool for
studies of magnetic fields. Namely, we would like to mention that
the change of the optical depth is an another important consequence of
atomic alignment. Such effect can be important for HI as was first
discussed by Varshalovich (1967). However, the actual calculations should
take into account the realignment caused by magnetic field.
It may happen that the variations of the optical depth caused by
alignment can be related to the Tiny Atomic Structures (TSAS) observed
in different phases of interstellar gas (see Heiles 1997).

\section{Summary}

In this paper we calculated alignment of various atomic species and
quantified the effect of magnetic fields on alignment. As the result,
we obtained the degrees of polarization that is expected for both scattering
and absorption.  
We have shown that

1. The existence of fine or hyperfine structure is a prerequisite for
atomic alignment.

2. The alignment of atoms happens as the result of interaction of atoms
with anisotropic flow of photons.

3. Atomic alignment can affect the polarization state of the scattered and
absorbed photons.

4. The degree of polarization is affected by mixing caused by Larmor
precession of atoms in external magnetic field. This allows a new
way to study magnetic field direction using polarimetry.

5. The degree of polarization depends on the angle of the magnetic
field with the anisotropic flow of photons.

6. Atomic alignment is an effect that is present for a number of
species. Combining them together allows  to get extra information
about magnetic field and environments in question.

7.  Time variations of magnetic field in interplanetary plasma should
result  in time variations of degree of polarization thus providing
a tool for interplanetary turbulence studies.

\begin{acknowledgments}
We thank Ken Nordsieck for fruitful discussions. And we acknowledge valuable commens from John mathis for. The work was supported
by the NSF grant  AST 0098597.

\end{acknowledgments}
  
\appendix
\section{Matrices}

Consider matrices corresponding to the projections of angular momentum
along Cartesian axes.

For instance, for $f=1/2,$

\begin{eqnarray}
F_{x}(0)=\frac{\hbar}{2}\left(\begin{array}{cc}
0& 1\\
1& 0\end{array}\right)\quad
F_{y}(0)=\frac{\hbar}{2}\left(\begin{array}{cc}
0 & -i\\
i & 0\end{array}\right)\quad
F_{z}(0)=\frac{\hbar}{2}\left(\begin{array}{cc}
1 & 0\\
0 & -1\end{array}\right)\nonumber, 
\end{eqnarray}
known as spin matrices.

For $f=3/2$,

\begin{eqnarray}
F_{x}(0)=\frac{\hbar}{2}\left(\begin{array}{cccc}
0 & \sqrt{3} & 0 & 0\\
\sqrt{3} & 0 & 2 & 0\\
0 & 2 & 0 & \sqrt{3}\\
0 & 0 & \sqrt{3} & 0\end{array}\right),\quad
F_{y}(0)=\frac{\hbar}{2}\left(\begin{array}{cccc}
0 & -\sqrt{3}i & 0 & 0\\
\sqrt{3}i & 0 & -2i & 0\\
0 & 2i & 0 & -\sqrt{3}i\\
0 & 0 & \sqrt{3}i & 0\end{array}\right),\quad
F_{z}(0)=\frac{\hbar}{2}\left(\begin{array}{cccc}
3 & 0 & 0 & 0\\
0 & 1 & 0 & 0\\
0 & 0 & -1 & 0\\
0 & 0 & 0 & -3\end{array}\right).
\end{eqnarray}
For $f=1$,

\begin{eqnarray}
F_{x}(0)=\frac{\hbar}{\sqrt 2}\left(\begin{array}{ccc}
0 & 1 & 0 \\
1 & 0 & 1\\
0 & 1 & 0\end{array}\right),\quad
F_{y}(0)=\frac{\hbar}{\sqrt 2}\left(\begin{array}{ccc}
0 & -i & 0\\
i & 0 & -i \\
0 & i & 0\end{array}\right),\quad
F_{z}(0)=\hbar\left(\begin{array}{ccc}
1 & 0 & 0\\
0 & 0 & 0\\
0 & 0 & -1\end{array}\right).
\end{eqnarray}
For f =2,
\begin{eqnarray}
F_{x}(0)=\frac{\hbar}{2}\left(\begin{array}{ccccc}
0 & 2 & 0 & 0 & 0\\
2 & 0 & \sqrt{6} & 0 & 0\\
0 & \sqrt{6} & 0 & \sqrt{6} & 0\\
0 & 0 & \sqrt{6} & 0 & 2\\
0 & 0 & 0 & 2 & 0\end{array}\right).\quad
F_{y}(0)=\frac{\hbar}{2}\left(\begin{array}{ccccc}
0 & -2i & 0 & 0 & 0\\
2i & 0 & -\sqrt{6}i & 0 & 0\\
0 & \sqrt{6}i & 0 & -\sqrt{6}i & 0\\
0 & 0 & \sqrt{6}i & 0 & -2i\\
0 & 0 & 0 & 2i & 0\end{array}\right).\quad
F_{z}(0)=\frac{\hbar}{2}\left(\begin{array}{ccccc}
2 & 0 & 0 & 0 & 0\\
0 & 1 & 0 & 0 & 0\\
0 & 0 & 0 & 0 & 0\\
0 & 0 & 0 & 1 & 0\\
0 & 0 & 0 & 0 & 2\end{array}\right).\end{eqnarray}

\end{document}